\documentclass[12pt,preprint]{aastex}

\begin{document}
\title{~~\\ ~~\\ Star Formation in H{\sc i} Selected Galaxies. II. HII Region Properties}
\author{J. F. Helmboldt\altaffilmark{1} \& R. A. M. Walterbos\altaffilmark{1}}
\email{helmbold@nmsu.edu, rwalterb@nmsu.edu}
\author{G. D. Bothun\altaffilmark{2}}
\email{nuts@bigmoo.uoregon.edu}
\author{K. O'Neil\altaffilmark{3}}
\email{koneil@gb.nrao.edu}

\altaffiltext{1}{Department of Astronomy, New Mexico State University, Dept. 4500 PO Box 30001 Las Cruces, NM 88003}
\altaffiltext{2}{Department of Physics, University of Oregon, Eugene, Oregon 97403}
\altaffiltext{3}{NRAO, P.O. Box 2 Green Bank, WV 24944 }

\received{?}
\accepted{?}


\begin{abstract}
A sample of 69 galaxies with radial velocities less than 2500 km s$^{-1}$ was selected from the H{\sc i} Parkes All Sky Survey (H{\sc i}PASS) to deduce details about star formation in nearby disk galaxies selected with no bias to optical surface brightness selection effects. Broad (B and R) and narrow band (H$\alpha$) images were obtained for all of these objects. More than half of the sample galaxies are late-type, dwarf disks (mostly Sc and Sm galaxies). We have measured the properties of the H{\sc ii} regions on H$\alpha$, continuum subtracted images, using the HII{\it phot} package developed by \citet{thi00}. All but one of the galaxies contained at least one detectable H{\sc ii} region. Examination of the properties of the H{\sc ii} regions in each galaxy revealed that the brightest regions in higher surface brightness galaxies tend to be more luminous than those in lower surface brightness galaxies.  A higher fraction (referred to as the diffuse fraction) of the H$\alpha$ emission from lower surface brightness galaxies comes from diffuse ionized gas (DIG).  H{\sc ii} region luminosity functions (LFs) co-added according to surface brightness show that the shapes of the LFs for the lowest surface brightness galaxies are different from those for typical spiral galaxies.  This discrepancy could be caused by the lowest surface brightness galaxies having somewhat episodic star formation or by them forming a relatively larger fraction of their stars outside of dense, massive molecular clouds.  In general, the results imply that the conditions under which star formation occurs in lower surface brightness galaxies are different than in more typical, higher surface brightness spiral galaxies.

\end{abstract}


\section{Introduction}

The extended nature of disk galaxies coupled with the brightness of the night sky makes the study of the properties of large samples of these objects difficult in optical wavelengths.  Any sample chosen from an optical catalog will inevitably be biased toward higher surface brightness objects, as the surface brightness of the night sky in the B-band, even at a dark site, is $\sim$23 mag arcsecond$^{-2}$.  The surface brightness bias can be avoided, however, by choosing a sample from an H{\sc i} survey since sky emission is virtually nonexistent at 21 cm.  Galaxies contained within such a catalog are certain to contain a significant amount of hydrogen gas and are therefore likely to be detected in H$\alpha$ emission.\par
Among the structures typically found in the ISM, H{\sc ii} regions are of particular interest as they trace well the amount of recent star formation and the initial star cluster mass function of a galaxy.  The number of luminous H{\sc ii} regions, the luminosity of the brightest H{\sc ii} regions \citep{ken88}, and the slope of the H{\sc ii} region luminosity function (LF) \citep{ken89, ban93, cal91} have all been found to depend slightly on morphological type.  However, these results come from optical studies; the effect of the exclusion of low surface brightness (LSB) disks is therefore not well constrained.\par
A sample of galaxies was chosen from the H{\sc i} Parkes All Sky Survey (H{\sc i}PASS) \citep{bar01} to be imaged in broad band B and R and narrow band H$\alpha$ to examine star formation properties free of the surface brightness bias. The overall properties of this sample are presented in paper I \citep{hel04}.  This paper will focus on the characteristics of the H{\sc ii} regions by presenting the results of automated H{\sc ii} region photometry.\par

\section{HII Region Photometry}
\subsection{The Sample}

A sample of 132 galaxies with declinations less than -65$^{\circ}$ and radial velocities less than 2500 km s$^{-1}$ was chosen from H{\sc i}PASS; due mainly to weather constraints, 68 of these 132 were imaged through broad-band B and R and narrow-band H$\alpha$ filters.  Because of extinction issues, lower priority was given during the observing run to galaxies that appeared nearly edge-on on their Digitized Sky Survey images (available through the NASA/IPAC Extra Galactic Database (NED)\footnote{NED is operated by the Jet Propulsion Laboratory, California Institute of Technology, under contract with the National Aeronautics and Space Administration.}).  We also include M83 as a nearby example of a typical spiral galaxy and obtained images in all three bands for it. There was no significant detection of integrated H$\alpha$ emission for only one of these galaxies.  Slightly more than half of the sample galaxies have morphological types that are Sd or later.  The sample contains a higher fraction of relatively blue, LSB galaxies than is typically seen in optical catalogs.  A comparison of observed and model values for the H$\alpha$ equivalent widths, star formation rates (SFRs) per unit H{\sc i} mass, and B-R colors shows that the star formation histories of these objects are best characterized by SFRs that began $\sim$4 Gyr ago and were higher in the past.  For a detailed description of the sample properties as well as the data acquisition, reduction, calibration, and measurements, see paper I.\par

\subsection{HII Region Fluxes and Diffuse Fractions}

Automated H{\sc ii} region photometry was performed on the H$\alpha$ continuum subtracted images utilizing the IDL program HII{\it phot} developed by \citet{thi00}.  The images were corrected for $[$NII$]$ contamination using the following relation derived from a linear least squares fit to the photometric and spectroscopic data of \citet{jan00}
\begin{equation}
\mbox{log }\frac{[NII]}{H\alpha}=(-0.057\pm0.0093)M_{R}+(0.78\pm0.069)(B-R)_{e}+(-2.62\pm0.17)
\end{equation}
where $M_{R}$ is the absolute R-band magnitude and $(B-R)_{e}$ is the color index measured within the half-light radius.  The scatter about this relation is $\sim$0.1 dex smaller than that used in paper I which did not utilize the galaxies' color indexes.  For our sample, the values for $[$NII$]$/H$\alpha$ computed with equation (1) range from about 5\%-6\% for the faintest, bluest galaxies to about 50\%-60\% for the brightest, reddest galaxies.\par
The basic procedure for the HII{\it phot} program is as follows.  On each image, a region of interest is selected around the galaxy which is convolved with a Gaussian kernel to remove any background fluctuations.  The region is then convolved with several other kernels of various sizes to identify possible H{\sc ii} regions which are then fit with models of different morphologies.  The models have the following functional form\par
\begin{equation}
I_{H\alpha}=I_{H\alpha,p}\mbox{ exp} \left [-\frac{(r-r_{o})^2}{2\sigma^2} \right ]
\label{eqn1}
\end{equation}
where $I_{H\alpha,p}$ is the peak H$\alpha$ surface brightness and the ratio of the inner ring radius, $r_{o}$, to the Gaussian width, $\sigma$, is varied among six different values, 0, 0.5, 1, 2, 4, and 8.  For each region, these six models are stretched and rotated to match the shape of the region; the model that produces the best fit is used to obtain values for the axis ratio, position angle, and {\sc fwhm} along the major and minor axes, {\sc fwhm}$_{maj}$ and {\sc fwhm}$_{min}$.\par
``Footprints'' of the H{\sc ii} regions are then constructed by altering the boundaries of the objects so that any pixels that were shared by more than one object become associated with the object that is best fit by its model morphology.  These model boundaries are then trimmed to create H{\sc ii} region ``seeds'' by rejecting any pixels that are less than half of the median value for the object.  These seeds are then grown iteratively by lowering the threshold for the isophotal boundary by 0.02 dex at a time until the slope of the surface brightness profile reaches some minimum or until no more acceptable pixels (i.e. pixels that do not belong to other seeds or pixels that are below some maximum value) are available.  Seeds that are best fit by their model morphologies are grown first.  For this paper, a limiting surface brightness slope of 1.5 EM pc$^{-1}$ was used in the growth process where EM is the emission measure\footnote{On each image, the H$\alpha$ surface brightness in each pixel was converted to emission measure assuming a temperature of 10$^{4}$K and case B recombination (i.e. EM=5$\times$10$^{17} \frac{I_{H\alpha}}{1 \mbox{\scriptsize{erg s}}^{-1} \mbox{\scriptsize{ cm}}^{-2} \mbox{\scriptsize{ }} \Box ''}$ cm$^{-6}$ pc)} in standard units of cm$^{-6}$ pc.  An example of this process is illustrated by the images of one of the sample galaxies displayed in Fig. 1.\par
After the H{\sc ii} region boundaries are determined, the flux within these boundaries is automatically measured.  However, these fluxes may contain a contribution from any diffuse ionized gas (DIG) in the galaxy along the line of sight.  HII{\it phot} estimates this contribution for each region using the flux from background pixels defined to be all of the pixels outside the region's boundary within a projected distance of 250 pc from its boundary.  Any background pixel that is at least 75\% surrounded by other background pixels within a circular region of 250 pc is then identified as a control point.  The median values for the circular regions surrounding the control points are then computed and a surface is fit to them to estimate the flux within the boundary of the H{\sc ii} region that is emitted by the DIG.  This is then subtracted off the total flux within the boundary to give a corrected flux that will be used in the construction of the H{\sc ii} region luminosity function (LF).  Following this, HII{\it phot} produces a "background" image where the pixels belonging to H{\sc ii} regions are replaced with the background surface fits.\par
For each galaxy, the flux from the galaxy on the background image was measured and compared to the total H$\alpha$ flux to determine the diffuse fraction, $f_{d}$, or the fraction of the total H$\alpha$ flux that originates from the DIG.  We also formally compute the error in each value of $f_{d}$ using the counts measured from the galaxy on the H$\alpha$ and R-band images, a measure of the uncertainty in the value used for the background subtraction for the H$\alpha$ continuum subtracted image, and an estimate of the uncertainty in the continuum subtraction.  The individual values for $f_{d}$ as well as their 1$\sigma$ errors are listed in Table 1.  It should be noted that for some of the galaxies, the errors in the diffuse fractions are large ($>$50\%).  The error in $f_{d}$ is influenced heavily by the uncerties in the continuum and background subtraction processes, causing the error for some of the more LSB galaxies and galaxies with relatively bright stellar continua to be larger.  In some cases, the counts from the DIG are low which also causes the relative error in $f_{d}$ to be high.\par

\begin{figure}
\plotone{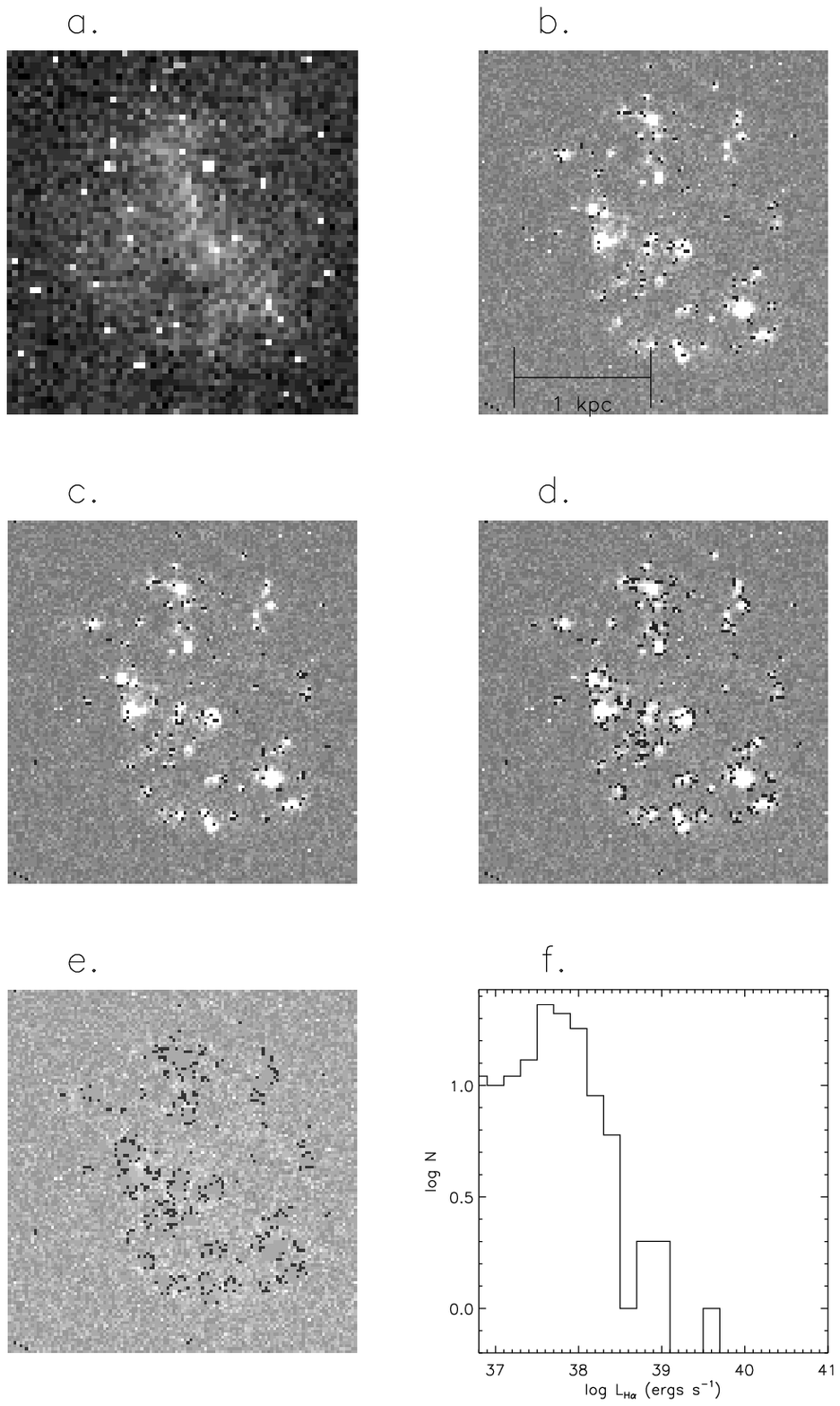}
\caption{Steps in the HIIphot procedure as performed on the Sm galaxy IC4710; (a) the B-band image, the H$\alpha$ continuum subtracted image with boundaries marked for (b) footprints, (c) seeds, and (d) final HII regions with the (e) diffuse background image and (f) the HII region LF.}
\label{fig1}
\end{figure}

\placefigure{fig1}
\placetable{tab1}

\subsection{The Brightest HII Regions}

For each galaxy, the mean H$\alpha$ luminosity and diameter of the three brightest H{\sc ii} regions were calculated; we will refer to these mean quantities hereafter as L$_{3}$ and D$_{3}$.  We chose to compute a weighted mean where the weights used for the two quantities for each H{\sc ii} region were $\left(\frac{\mbox{\sc snr}}{L}\right)^{2}$ and $\left(\frac{\mbox{\sc snr}}{D}\right)^{2}$ respectively where {\sc snr} is the signal to noise ratio computed by HII{\it phot}.  This weighting scheme was chosen because many galaxies contained few if any large H{\sc ii} regions implying that the {\sc snr} for any one of the three brightest regions may be relatively low.  For each H{\sc ii} region, the diameter was taken to be the diameter of a circle that occupies the same area on the image as the H{\sc ii} region.  To correct for the effects of the {\sc psf}, we first measure the median {\sc fwhm} of Gaussian fits to the radial profiles of ten stars on the H$\alpha$ image of each galaxy.  We then used the effective circular {\sc fwhm} from the Gaussian model fit for each region (equal to $\sqrt{\mbox{\sc fwhm}_{maj} \; \mbox{\sc fwhm}_{min}}$) to effectively deconvolve the measured diameter for each region according to
\begin{equation}
\mbox{D}_{corr} = \mbox{D } \frac{\sqrt{\mbox{\sc fwhm}_{mod}^{2} - \mbox{\sc fwhm}_{psf}^{2}}}{\mbox{\sc fwhm}_{mod}}
\end{equation}
were {\sc fwhm}$_{mod}$ is the effective circular {\sc fwhm} from the Gaussian model fit.  The weighted mean diameter for the three brightest H{\sc ii} regions computed using these corrected diameters is referred to as D$_{3,corr}$ and is listed for each galaxy in Table 1.\par
To compare our results with those of \citet{ken88}, we plot in Fig. 2 L$_{3}$ as a function of total B-band absolute magnitude corrected for internal extinction according to the correction used to produce the $B_{T}^{0}$ magnitudes in the Revised Shapley-Ames Catalog \citep{san81} used by \citet{ken88}, which we refer to as $M_{B}(0)$.  We fit a line to this data for Sbc/Sc galaxies only and find that the trend for our galaxies is somewhat less steep than the fit obtained for the Sbc/Sc galaxies from \citet{ken88}; both lines are plotted with the data in Fig. 2.  We also plot the fit to the Sbc/Sc galaxies in our sample $\pm 1 \sigma$ where $\sigma$ is the rms scatter about the fitted line.  These lines demonstrate that while the slope for the fit to our data is formally less steep, the two fits agree within 1$\sigma$.  We find, as \citet{ken88} did, that a significant number of Sm and Im galaxies in our sample lie above the line fit to the Sbc/Sc galaxies.  We also find that a small but significant fraction of these extremely late-type disk galaxies lie near or below both the line fit to our data and the line fit to the data of \citet{ken88}.  In contrast, all but one of the Sm and Im galaxies within the sample of \citet{ken88} lie above the fit to the data for the Sbc/Sc galaxies.  It should be noted that \citet{ken88} used isophotal H{\sc ii} region luminosities and computed the average of the three brightest regions without using any weighting.  To examine the effect of this on our measured trend, we also computed the luminosity of the H{\sc ii} regions for each galaxy using the same limiting isophote as \citet{ken88}, 100 cm$^{-6}$ pc.  The median difference between the value for L$_{3}$ and the unweighted mean of the isophotal luminosities for the three brightest H{\sc ii} regions is 0.07 dex, implying that the difference between the two techniques used for measuring luminosities contributes weakly, if at all, to the difference in the two observed trends.\par

\begin{figure}
\plotone{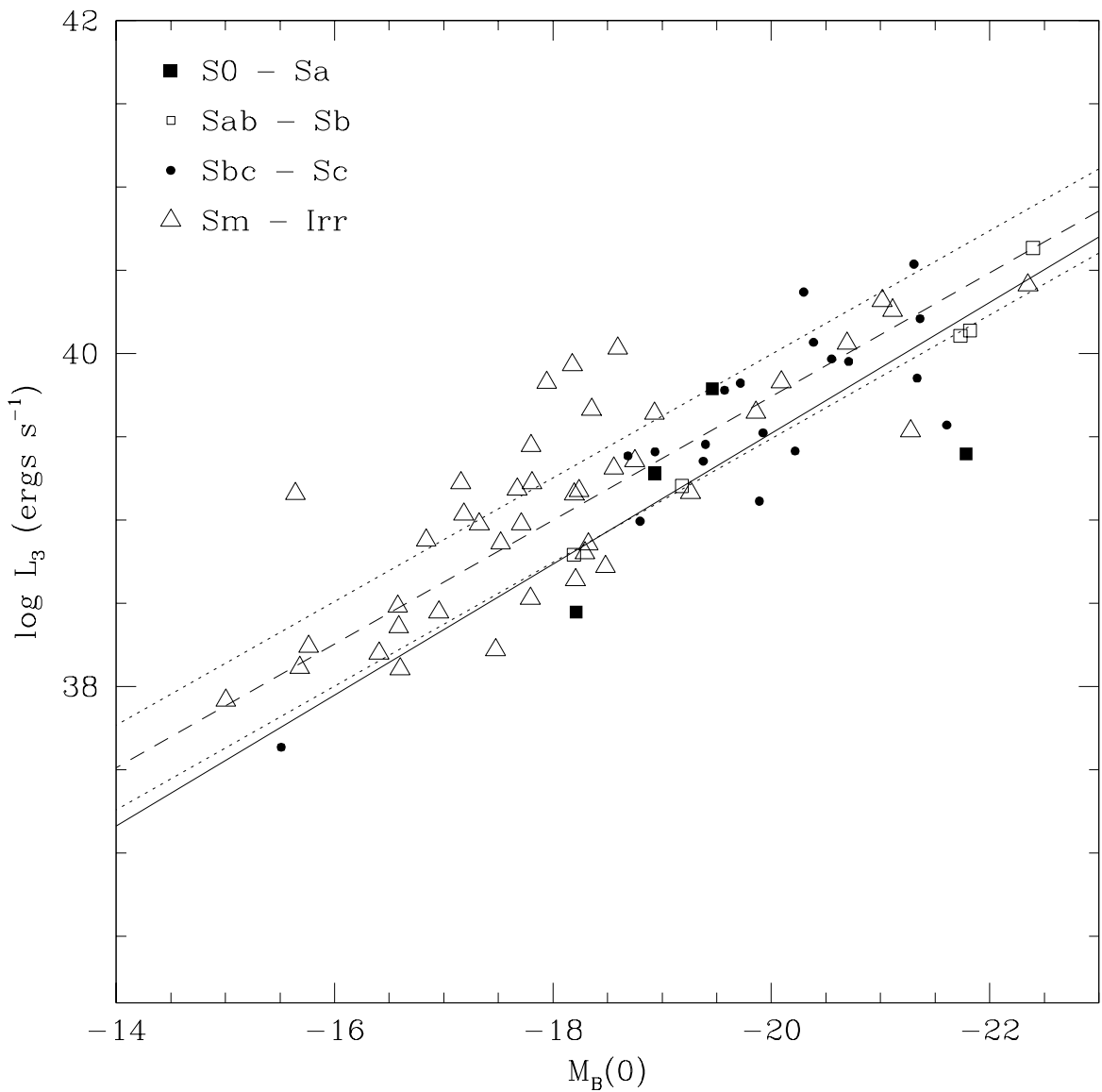}
\caption{Absolute B-band magnitude corrected for internal extinction according to \citet{san81} versus the weighted mean luminosity of the three brightest H{\sc ii} regions, L$_{3}$; the dashed line is a linear least squares fit to the data for Sbc/Sc galaxies only; the dotted lines are this fit $\pm 1 \sigma$; the solid line line represents the fit applied to data for Sbc/Sc galaxies taken from Fig. 3 of \citet{ken88}.}
\label{fig2}
\end{figure}

\placefigure{fig2}
  
\subsection{Measured Properties and Biases}

For H$_{\circ}$=70 km s$^{-1}$ Mpc$^{-1}$, one arcsecond corresponds to about 170 pc at the radial velocity limit of our sample.  This implies that the ability of HII{\it phot} to find smaller, lower luminosity H{\sc ii} regions is limited for the most distant galaxies in the sample.  To explore the effect this has on the measured values of L$_{3}$, D$_{3}$, and $f_{d}$, we first compute a minimum detectable H{\sc ii} region H$\alpha$ luminosity, L$_{lim}$, for each galaxy.  We do this by fitting a line to log L$_{H\alpha}$ as a function of log {\sc snr} for all the regions in the galaxy.  This fit was used to solve for the luminosity at a signal to noise ratio of five, the H{\sc ii} region detection limit recommended by \citet{thi00}.  The value for L$_{lim}$ was then taken to be this luminosity plus 1$\sigma$, where $\sigma$ is the rms deviation of the log L$_{H\alpha}$ values from the fitted line.  All but five of the sample galaxies have L$_{lim} > 10^{38}$ ergs s$^{-1}$ cm$^{-2}$; the largest value is 10$^{38.5}$ ergs s$^{-1}$ cm$^{-2}$.  Because of this, we list in Table 1 the number of regions found by HII{\it phot} with log L$_{H\alpha} >$38 rather than the total number of regions detected so that a comparison of the number of regions found in one galaxy to that found in another can be made without any bias caused by the completeness limit increasing with distance.\par
In Fig. 3, we plot radial velocity relative to the Local Group and corrected for in-fall into Virgo, $V_{LG}$, versus L$_{lim}$.  As expected, L$_{lim}$ increases with $V_{LG}$ but with a significant amount of scatter ($\sim$0.3 dex about a line fit to the data).  In the lower panel of Fig. 3, we plot $V_{LG}$ versus L$_{3}$.  We also plot the median value of L$_{3}$ in five bins for V$_{LG}$; we display these values as shaded rectangles with widths equal to the widths of the V$_{LG}$ bins and with lower and upper boundaries that correspond to the lower and upper quartiles for L$_{3}$ for the bins.  It is evident from this plot that there is a bias toward higher values of L$_{3}$ for the more distant galaxies.  However, the largest values of L$_{3}$ for the most distant galaxies are not significantly higher than the largest values for the most nearby galaxies, implying that this bias is due mostly to the increase in typical detection limit with $V_{LG}$ than to a problem with artificial blending of luminous (log L$_{H\alpha} \gtrsim$38) H{\sc ii} regions for more distant objects.  This is consistent with tests run by \citet{thi00} on M51 which implied that blending had no significant effect on the number of H{\sc ii} regions with log L$_{H\alpha} \geq$38.6 detected by HII{\it phot} for distances up to 45 Mpc \citep[see Fig. 11 of][]{thi00}.\par
The same is not true, however, for D$_{3}$.  We plot $V_{LG}$ versus D$_{3}$ in Fig. 4 along with median values for five $V_{LG}$ bins and find that there is an apparent positive correlation between the two quantities.  This is what one may expect if artificial blending of H{\sc ii} regions were a significant problem at higher redshift.  However, the absence of a similar trend between $V_{LG}$ and L$_{3}$ indicates that this is most likely not the case.  It is then more likely that this trend is caused by the increase in the physical size that corresponds to the size of the seeing disk for the more distant galaxies.  The median {\sc fwhm} of the {\sc psf} for our H$\alpha$ images was about 1{\huge \H{.}}6, or about 280 pc for $V_{LG}$=2500 km s$^{-1}$.  To estimate the magnitude of the effect the size of the seeing disk has on the trend in Fig. 4, we plot D$_{3,corr}$ as a function of $V_{LG}$ in the lower panel of Fig. 4.  From this plot, it can be seen that the {\sc psf} corrections from equation (3) significantly reduce the magnitude of the trend seen in the upper panel of Fig. 4 while retaining a bias toward larger values of D$_{3}$ at larger distances similar to the bias observed in Fig. 3 for L$_{3}$.  This implies again that blending of H{\sc ii} regions is not a significant problem for our sample for the brightest regions.  It also implies that the measured H{\sc ii} region sizes are significantly affected by the width of the {\sc psf} and are not reliable indicators for how the typical region diameter may depend on various galaxy properties.\par
We plot $f_{d}$ as a function of $V_{LG}$ and median values for $f_{d}$ in five $V_{LG}$ bins in Fig. 5 to explore the possibility that the higher detection limit at larger distances causes the amount of DIG to be overestimated.  If this is the case, one would expect the diffuse fractions to be significantly higher at larger redshifts.  From Fig. 5, it is evident that the median value for $f_{d}$ for the lowest $V_{LG}$ bin is smaller than that for the other bins.  However, the values for the remaining four bins are roughly constant, implying that there is not a significant bias toward higher values of $f_{d}$ at larger distances.

\begin{figure}
\plotone{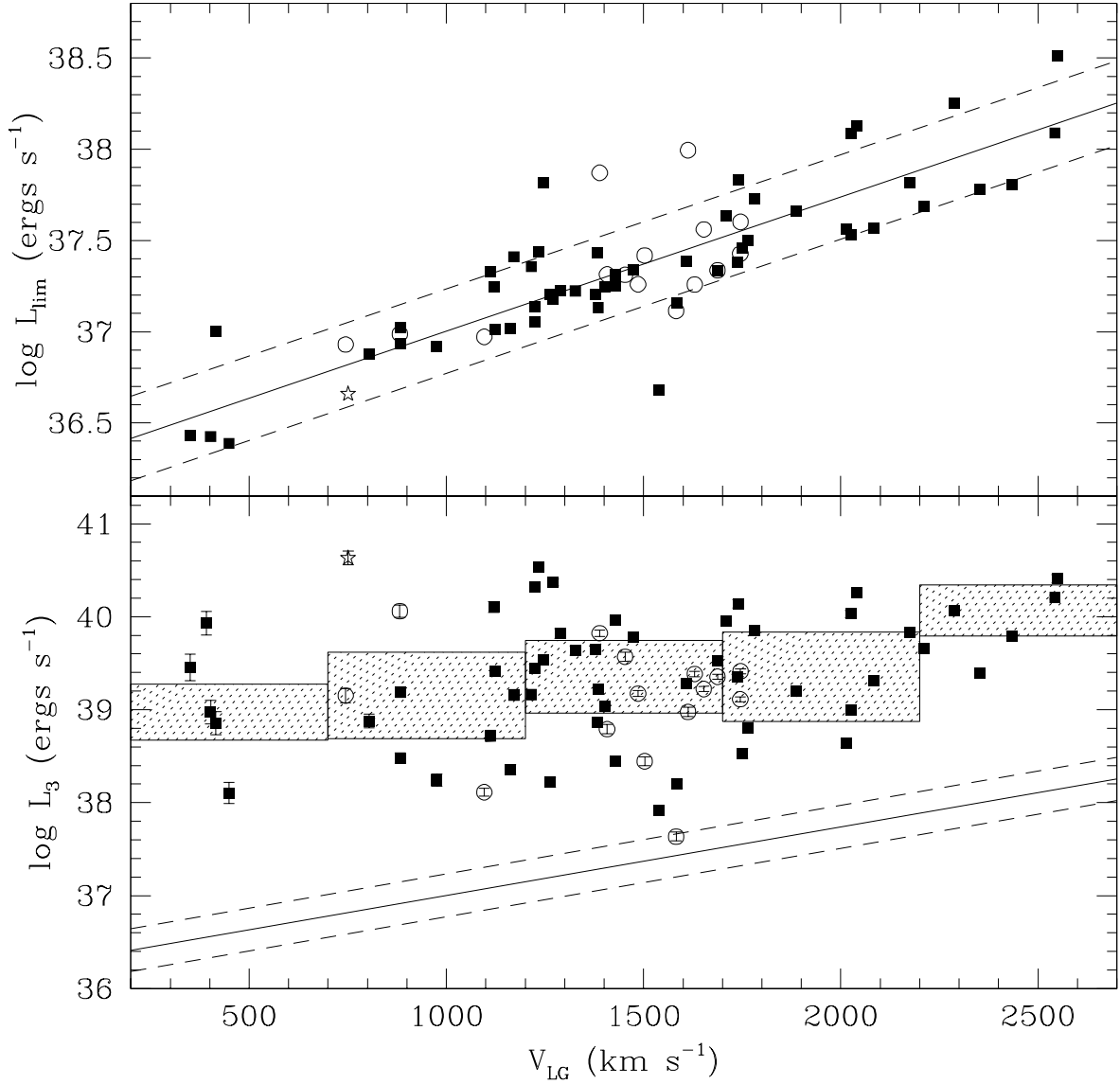}
\caption{Limiting H{\sc ii} region H$\alpha$ luminosity (i.e. the luminosity at which {\sc snr}$\approx$5), L$_{lim}$, for each galaxy as a function of radial velocity relative to the Local Group corrected for infall into Virgo, $V_{LG}$, (upper) with a linear least squares fit to the data (solid line) $\pm 1 \sigma$ (dashed lines).  The weighted mean luminosity for the three brightest H{\sc ii} regions in each galaxy, L$_{3}$, as a function of $V_{LG}$ (lower) with the fit to log L$_{lim}$ as a function of $V_{LG}$ $\pm 1 \sigma$ plotted as solid and dashed lines.  The median values for L$_{3}$ for five $V_{LG}$ bins are plotted as shaded boxes where the upper and lower boundaries of the boxes correspond to the upper and lower quartiles for the bins.  Open circles represent galaxies where the calibration is questionable (see paper I); the star is M83.}
\label{fig3}
\end{figure}

\begin{figure}
\plotone{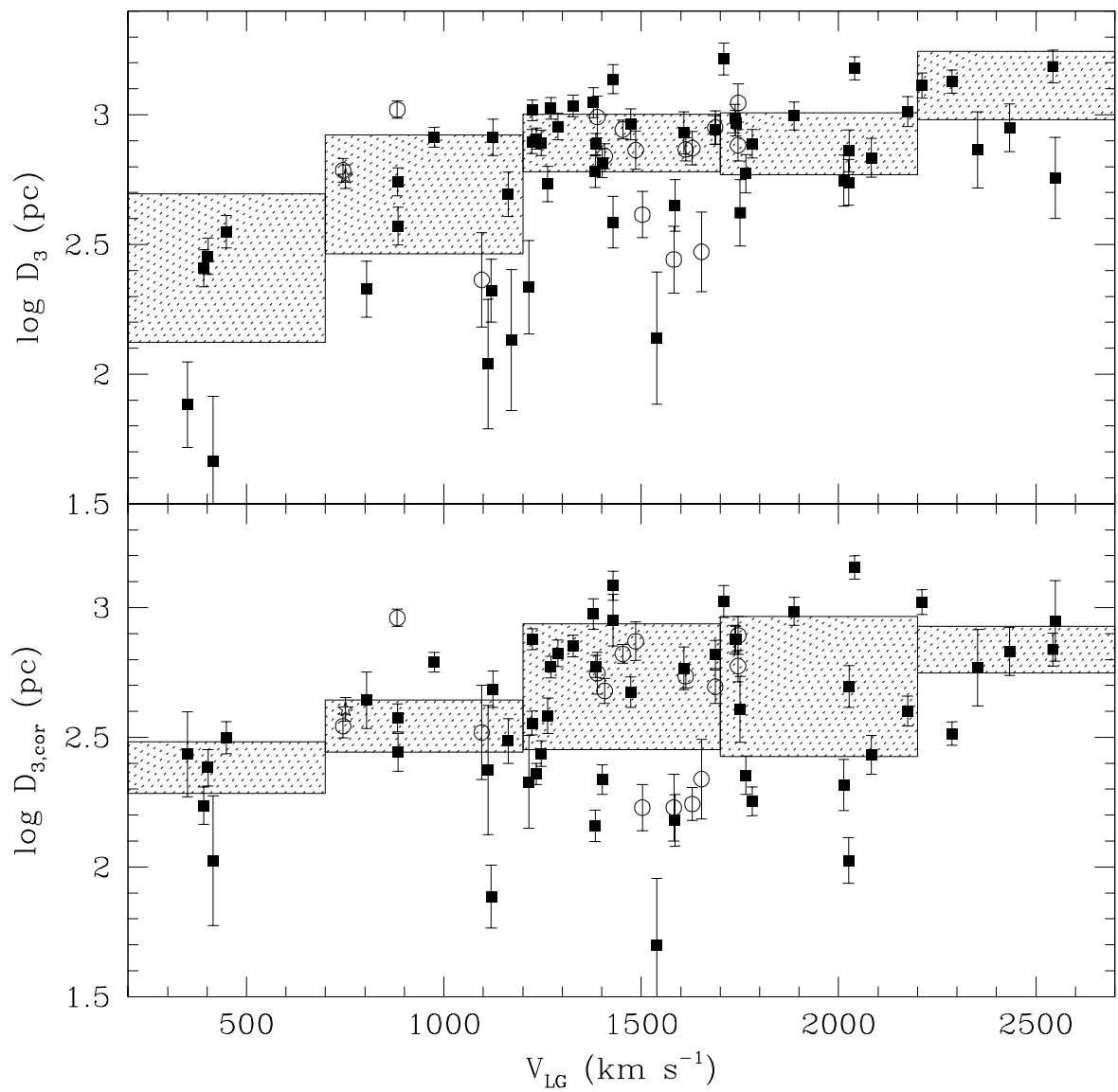}
\caption{Weighted mean diameter of the three brightest H{\sc ii} regions, D$_{3}$, as a function of radial velocity, $V_{LG}$; the diameters in the lower panel are the values including a correction for a Gaussian {\sc psf} according to equation (3).  The symbols and shaded boxes have the same meanings as in Fig. 3.}
\label{fig4}
\end{figure}

\begin{figure}
\plotone{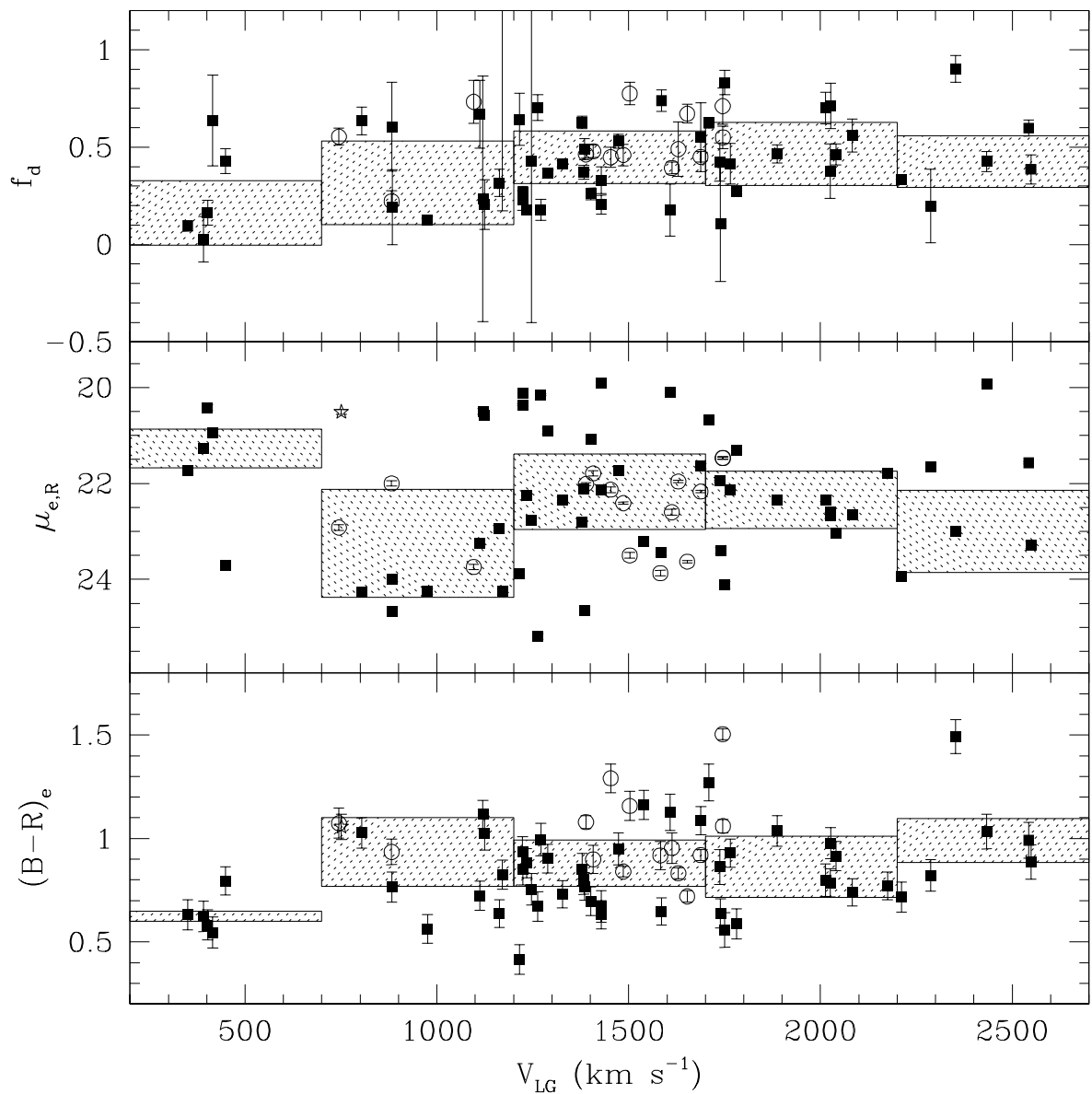}
\caption{The diffuse fraction, $f_{d}$, R-band surface brightness at the half-light radius, $\mu_{e,R}$, and color within the half-light radius $(B-R)_{e}$, as functions of $V_{LG}$.  The symbols and shaded boxes have the same meanings as in Fig. 3.}
\label{fig5}
\end{figure}

\placefigure{fig3}
\placefigure{fig4}
\placefigure{fig5}

\subsection{Surface Brightness and Color}

In paper I, it was discovered that the H{\sc i}PASS sample contains a larger fraction of galaxies that are relatively bluer and lower surface brightness than the optically selected Nearby Field Galaxy Survey (NFGS) of \citet{jan00}.  To explore the possibility that the properties of the H{\sc ii} regions and DIG in the bluest and most LSB galaxies which appear in smaller numbers in similar optical samples may be significantly different, we first plot the R-band surface brightness at the half-light radius, $\mu_{e,R}$, and the color within the half-light radius, $(B-R)_{e}$, as functions of $V_{LG}$ in Fig. 5.  We again plot median values within five $V_{LG}$ bins and display the results as shaded rectangles in the same way as was done for L$_{3}$ in Fig. 3 (see \S 2.4).  From these plots, there appears to be no significant bias toward higher or lower surface brightness galaxies at higher redshift.  Similarly, there is no apparent bias toward bluer or redder galaxies at larger distances.  Because of this, any trend observed between L$_{3}$ and either surface brightness or color cannot be the result of the bias toward larger values of L$_{3}$ at higher redshift that is evident in Fig. 3.  To explore possible relationships among these quantities, we plot $\mu_{e,R}$ and $(B-R)_{e}$ versus L$_{3}$ and $f_{d}$ for all of our sample galaxies in Fig. 6.  We do not include similar plots using D$_{3}$ because of inaccuracies caused by the relative size of the seeing disk as described in the previous section.\par
Within four bins in surface brightness and color, each containing roughly the same number of galaxies, we compute the median values for L$_{3}$ and $f_{d}$ and display them in the appropriate plots in Fig. 6 as shaded rectangles with widths equal to the width of the bins in $\mu_{e,R}$ and $(B-R)_{e}$ and with upper and lower boundaries that correspond to the median values for L$_{3}$ and $f_{d}$ $\pm 1 \sigma$ where $\sigma$ is the error in the median value (i.e. not the difference between the upper and lower quartiles as was used in previous figures).  We also plot median values for L$_{3}$ and $f_{d}$ for all galaxies as horizontal dashed lines.  From these median values, it can be seen that the lowest surface brightness galaxies have higher diffuse fractions and lower values for L$_{3}$; the same is true to a lesser degree for relatively bluer galaxies.  The fact that lower surface brightness galaxies are found to have lower values of L$_{3}$ may be heavily influenced by small number statistics as the number of detected H{\sc ii} regions is correlated with surface brightness.  The degree to which this may effect the observed trend will be addressed later in \S 3.4.  The trend found between diffuse fraction and surface brightness is similar to the trend found by \citet{gre98} who demonstrated that within individual galaxies, areas with higher SFR surface densities had lower diffuse fractions.  Despite the instances of highly uncertain diffuse fractions alluded to in \S 2.2, the errors in the median values for the four surface brightness bins are small enough to conclude that the trend observed between $\mu_{e,R}$ and $f_{d}$ is real.

\begin{figure}
\plotone{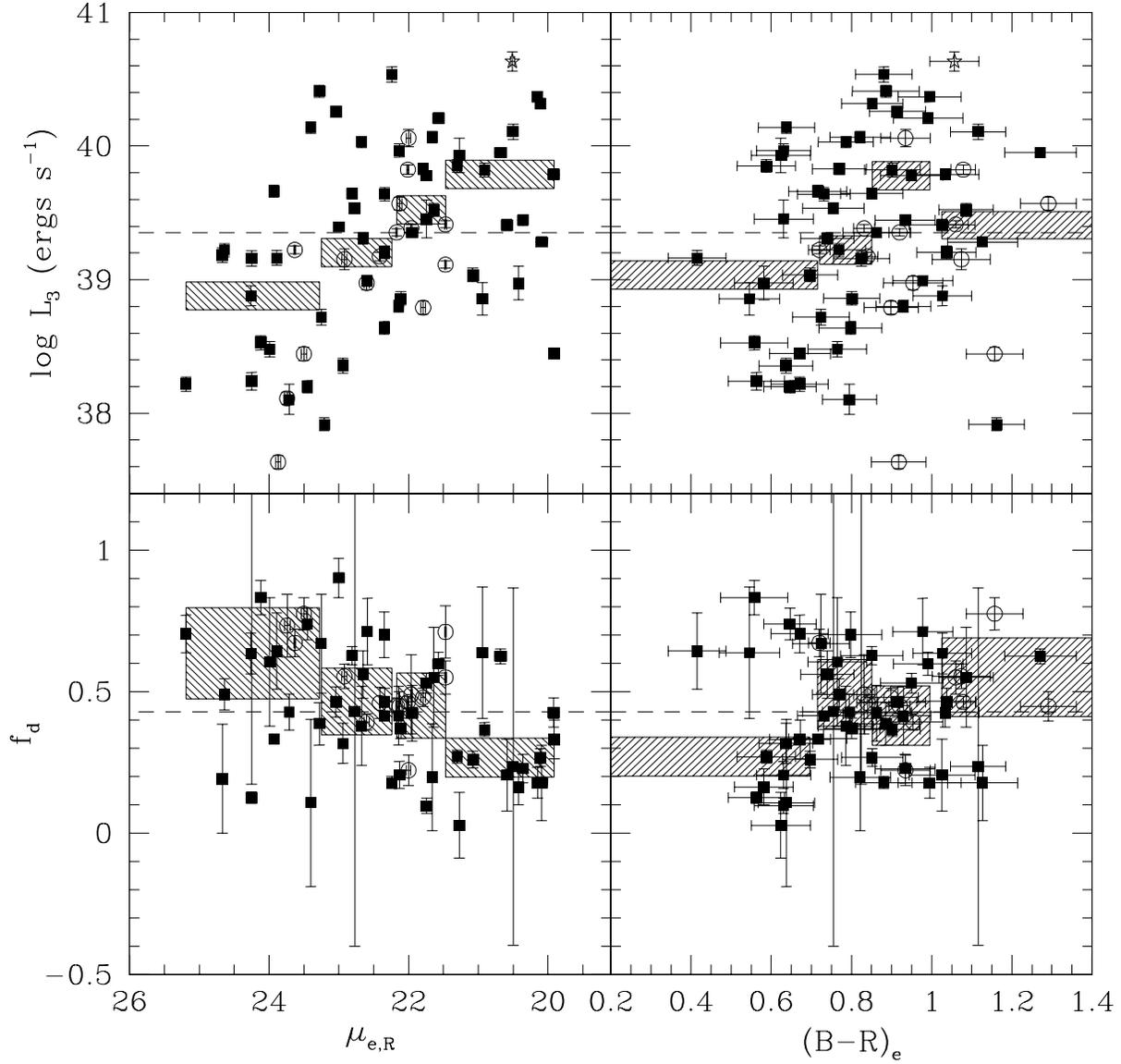}
\caption{Weighted mean luminosity of the three brightest HII regions, L$_{3}$, and diffuse fraction, $f_{d}$, as functions of surface brightness at the half-light radius, $\mu_{e,R}$, and color within the half-light radius, $(B-R)_{e}$.  The symbols have the same meaning as in Fig. 3.  The median values for L$_{3}$ and $f_{d}$ in four bins for $\mu_{e,R}$ and $(B-R)_{e}$ are displayed as shaded boxes in the appropriate plots; the width of the boxes correspond to the width of the bins; the upper and lower limits of the boxes correspond to the median values $\pm 1 \sigma$ where $\sigma$ is the error in the median value.  The median values for L$_{3}$ and $f_{d}$ for all galaxies are displayed as dashed lines.}
\label{fig6}
\end{figure}

\placefigure{fig6}

\section{HII Region LFs}

\subsection{Individual LFs}
H{\sc ii} region LFs were constructed for all galaxies within our H{\sc i}PASS sample with $V_{LG}>$840 km s$^{-1}$.  This velocity requirement was used because the major sources of uncertainty in the corrected velocities come from the errors in the assumed values for the velocity of the sun relative to the Local Group and the distance to the Virgo cluster.  These uncertainties produce errors in the corrected radial velocities that are on the order of 50 km s$^{-1}$ \citep{yah77} and 80 km s$^{-1}$ \citep{lu94} respectively yielding a total uncertainty of about 94 km s$^{-1}$.  Given that the luminosity bins for the LFs were chosen to be 0.2 dex, the error in the luminosity originating from the error in the distance determination must be less than 0.1 dex for the computed LF to accurately reflect the true LF for a given galaxy.  Therefore, the largest allowed error in $V_{LG}$ is 11.5\%, and the minimum allowable corrected radial velocity is 817.4 km s$^{-1}$.  A more conservative limit of 840 km s$^{-1}$ was used to ensure that the shapes of the LFs would not be significantly affected by distance errors and because this velocity corresponds to a distance of 12 Mpc for H$_{\circ}$=70 km s$^{-1}$ Mpc$^{-1}$.  For each galaxy, only H{\sc ii} regions with signal to noise ratios greater than five were considered to be detected regions \citep[i.e. the detection limit recommended by][]{thi00}.  Example LFs are plotted in Fig. 7.\par
Previous studies of H{\sc ii} region LFs \citep[e.g.][]{ham01} have used power law fits to the bright end of the LFs to characterize their shapes, assuming the LF follows the form dN$\propto$L$^{\alpha}$dL.  Fits are typically confined to the largest values of L$_{H \alpha}$ because for a \citet{sal55} initial mass function (IMF) (i.e. power law slope of 2.35) and an upper limit of 100 M$_{\odot}$, the largest H$\alpha$ luminosity one would expect from a region ionized by a single star is about 3$\times 10^{38}$ ergs s$^{-1}$.  \citet{oey98} demonstrated that for a distribution of regions ionized by poor or "unsaturated" clusters, the LF drops steeply at an H$\alpha$ luminosity of about 10$^{39}$ ergs s$^{-1}$; beyond this, the shape of the LF traces well the distribution of masses for rich clusters.  Because of this, for any galaxy that forms a significant number of poor clusters, the shape of the LF can be markedly different below and above the limit of 10$^{39}$ ergs s$^{-1}$.  Therefore, to characterize the shapes of the LFs, we fit a power law for log L$_{H\alpha}>$39 for all LFs where at least five luminosity bins with log L$_{H\alpha}>$39 are not empty.  The results of these fits are plotted with the appropriate LFs in Fig. 7; values for the power law slope, $\alpha$, for individual galaxies are listed in Table 1.  No significant trend was found between $\alpha$ and either $\mu_{e}$ or $(B-R)_{e}$.  It is possible that this is due to the large uncertainty in the determination of $\alpha$ (the median error is $\sim$20\%) or to the fact that more LSB, fainter galaxies tend to form few if any regions with log L$_{H\alpha}>$39, implying that these power law fits do not provide any information regarding the difference between the shapes of the LFs for these galaxies and those for more HSB, brighter galaxies.\par

\begin{figure}
\plotone{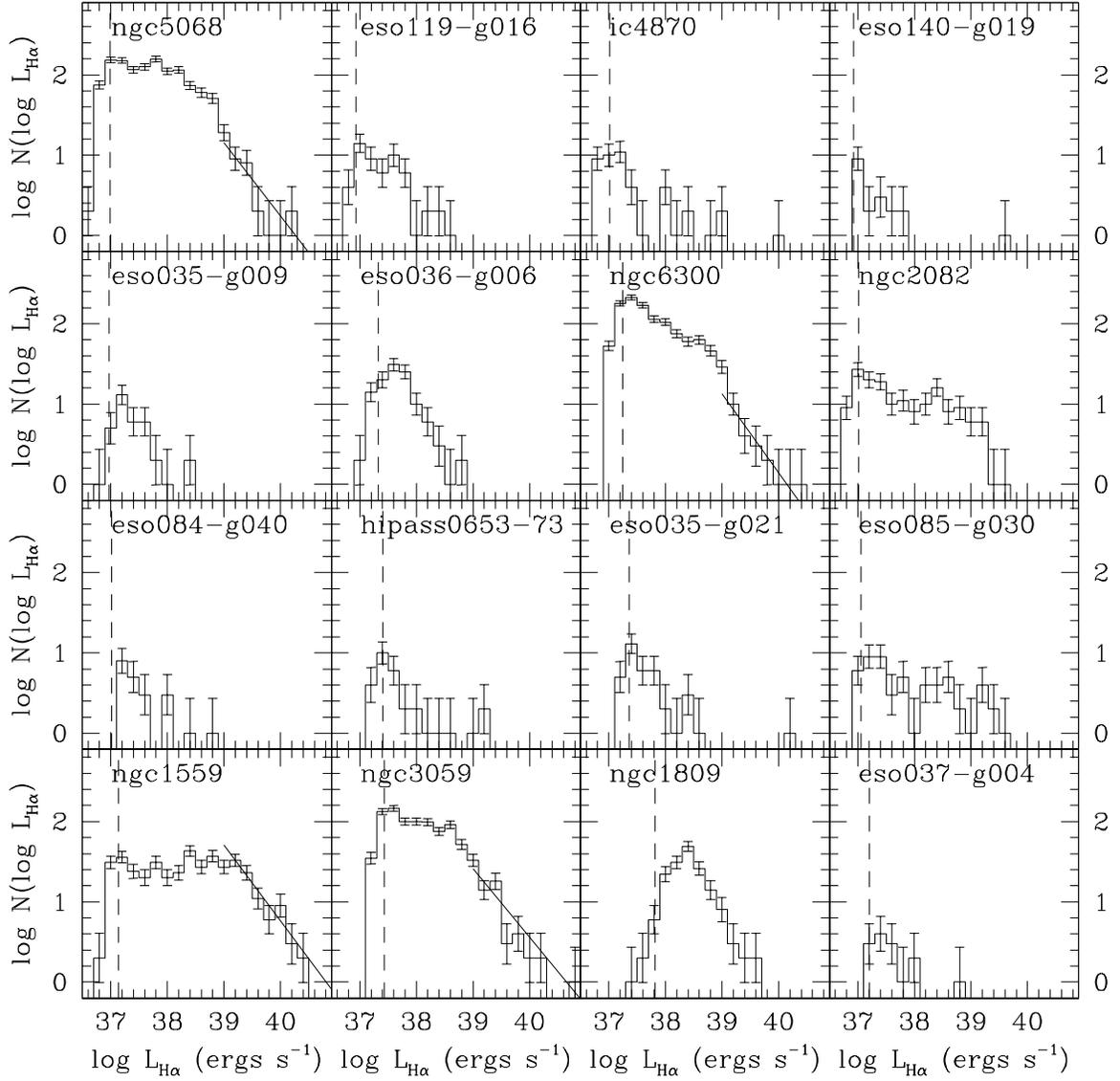}
\caption{HII region LFs for the 16 most nearby galaxies with V$_{LG}>$840 km s$^{-1}$; weighted linear least squares fits for log L$_{H\alpha}>$39 are displayed for galaxies where at least five of the luminosity bins with log L$_{H\alpha}>$39 are not empty; the vertical dashed lines represent the completeness limit, L$_{lim}$, for the galaxies' LFs (see \S 2.4).}
\label{fig7}
\end{figure}

\placefigure{fig7}

\subsection{Co-added LFs}
To explore how the shapes of the H{\sc ii} region LFs for the lowest surface brightness and bluest galaxies in the H{\sc i}PASS sample may differ from those for galaxies that are more typically contained within optically selected samples such as the NFGS, we have constructed co-added H{\sc ii} region LFs.  We have opted for this approach because, as stated above, the most LSB and bluest galaxies tend to be less luminous and to have fewer detectable H{\sc ii} regions.  The co-adding process used was as follows.  For a particular co-added group, within each luminosity bin, a weighted mean number of regions was computed for all galaxies that had L$_{lim}$ (see \S 2.4) less than the lower boundary of the bin.  Since larger galaxies tend to form more H{\sc ii} regions and since the H{\sc i}PASS sample was selected from an H{\sc i} catalog, for this computation, the number of regions within each bin from each galaxy was weighted by $M_{HI}^{-1}$.  This was done to ensure that no one galaxy dominated the shape of the resulting co-added LF.  Because of this weighting scheme, only galaxies with $M_{HI}$ greater than the H{\sc i}PASS 3$\sigma$ detection limit at a radial velocity of 2500 km s$^{-1}$ (1.6$\times10^{8}$ M$_{\odot}$ for H$_{\circ}$=70 km s$^{-1}$ Mpc$^{-1}$) were included.  Again, galaxies with corrected radial velocities less than 840 km s$^{-1}$ were excluded.  These two requirements qualify 58 of the 69 H{\sc i}PASS galaxies for the co-adding process.\par
To asses how distance effects influence this co-adding process, all galaxies eligible to be co-added were sorted by radial velocity and then placed in three roughly equal sized groups designated near, intermediate, and far.  The ranges in values of $V_{LG}$ for the three groups are roughly 880$<V_{LG}<$1380 km s$^{-1}$ for the near group, 1380$<V_{LG}<$1730 km s$^{-1}$ for the intermediate group, and 1730$<V_{LG}<$2500 km s$^{-1}$ for the far group.  The co-added LFs for these groups are displayed in Fig. 8 along with the ratio of the near group LF to the intermediate group LF and the ratio of the near group LF to the far group LF as functions of luminosity.  For log L$_{H\alpha}>$38, the LF for the intermediate group appears to have fewer regions at all luminosities than the nearby group; the opposite is true for the LF for the far group.  However, the residuals between the near group LF and both the intermediate and far group LFs are relatively flat as functions of H$\alpha$ luminosity for 38$\leq \mbox{log } L_{H\alpha} \leq$40, implying that the shape of the co-added LFs do not change significantly with distance.\par
\citet{thi00} used images of M51 altered to make the galaxy appear more distant to asses the effects of sensitivity and blending on the shape of the LF as a function of distance.  The altered images were created to simulate the appearance of M51 if it were at distances of 15, 30, and 45 Mpc (the actual distance to M51 was assumed to be 9.6 Mpc).  The results from this analysis predict that for more distant galaxies, HII{\it phot} will measure an LF with a slope that is artificially flattened by blending for intermediate luminosity regions while maintaining the same shape for higher luminosity regions (log L$_{H\alpha} \geq$38.6 for the case of M51 for a distance of up to 45 Mpc).  Since the most distant galaxies in our sample are closer than 40 Mpc and the shapes of the LFs for all three distance groups agree reasonably well, flattening of the LFs from distance effects does not appear to be a significant problem for our co-adding process.\par
While these results imply that the effects of blending do not depend on distance within the volume occupied by our sample, they may be worse for locations where H{\sc ii} regions are more closely spaced \citep[e.g.][]{sco01} such as spiral arms or higher surface brightness galaxy disks.  This may cause the shapes of the measured LFs for such locations to artificially appear less steep when compared to LFs for locations where the typical separation between H{\sc ii} regions is larger such as interarm regions and LSB galaxies.  The possible impact of this on the analysis to follow will be discussed in \S 3.3.\par
To examine how the LFs for the most LSB and bluest galaxies may differ from those for typical spirals, six co-added groups were selected.  First, LFs for two comparison groups were constructed.  The comparison groups were chosen so that they spanned the ranges in surface brightness and color typical of optical catalogs and so that they contained enough galaxies that the resulting LFs would be well determined (i.e. less than 10\% error for the majority of the log L$_{H\alpha}$ bins).  The surface brightness comparison group was chosen to consist of all galaxies from our sample that qualified to be co-added according to the criteria described above that were within $\pm 2 \sigma$ of the mean surface brightness for the NFGS (24.4$<\mu_{e,B}<$20.4), 50 galaxies in all.  The color comparison group was chosen in a similar way by selecting all qualified galaxies from our sample within $\pm 2 \sigma$ of the mean color for the NFGS (0.60$<$(B-R)$_{e}<$1.60), resulting in a group of 56 galaxies.\par
To examine how the shapes of the LFs for galaxies that are at the extreme ends of the color and surface brightness distributions may differ from the typical shapes represented by the LFs computed for the comparison groups, the remaining four co-added groups were constructed as follows.  Among the 58 galaxies that qualify to be co-added, the ten lowest and ten highest surface brightness galaxies were co-added in two separate groups; similarly, the ten bluest and ten reddest galaxies were co-added in two other groups.  These groups are hereafter referred to as the LSB, HSB, blue and red groups respectively.  The number of galaxies within these groups was chosen to be ten as that was the number required to keep the errors within the majority of the log L$_{H\alpha}$ bins to less than 30\% for the LFs for all four groups.  The resulting co-added LFs for all six groups are plotted in the two left panels of Fig. 9.\par
At the lowest detectable H{\sc ii} region luminosities, all the LFs appear to roughly agree.  However, the shape of the LSB group LF is markedly different from that of the comparison LF; the LSB group LF is approximated well by a power law down to log L$_{H\alpha}\sim$38.  The slope of the LF for the blue group appears to be flatter than that for the comparison and red groups for intermediate luminosities (37.5$\leq$log L$_{H\alpha} \leq$38.5); the slope for the red group LF appears to be steeper than that of the color comparison LF.  To quantify these differences in shape, we compute the reduced $\chi^{2}$ (referred to hereafter as $\chi^{2}_{red}$) between the LSB, HSB, blue, and red group LFs and their comparison LFs.  We do this by assuming that the LFs for the four extreme groups each differ from the corresponding comparison LF by a scale factor.  We then determine the scale factor that minimizes $\chi^{2}$ over the range 37.7$\leq$log L$_{H\alpha}<$40.5 and divide the minimized $\chi^{2}$ by the number of nonempty log L$_{H\alpha}$ bins over this range minus one.  This lower limit for the range in log L$_{H\alpha}$ was chosen because less than half of the galaxies in the comparison groups have log L$_{lim}<$37.7, implying that below log L$_{H\alpha}$=37.7, the completeness corrections made during the construction of the comparison LFs are significant and that this portion of each LF is not as representative of the entire co-added group as the rest of the LF.  The upper limit was chosen because beyond log L$_{H\alpha}$=40.5, only one or two galaxy contribute to each LF as evident by the large errorbars seen in Fig. 9.  The resulting values for $\chi^{2}_{red}$ are printed in the appropriate panels in Fig. 9.  From these values, it can be seen that the shapes of the HSB and blue group LFs agree with those of the their comparison LFs within roughly 1$\sigma$.  The shapes of the LSB and red group LFs do not match those of their comparison LFs as well with $\chi^{2}_{red}$ being about 2.5 and 1.7 for the LSB and red groups respectively.

\begin{figure}
\plotone{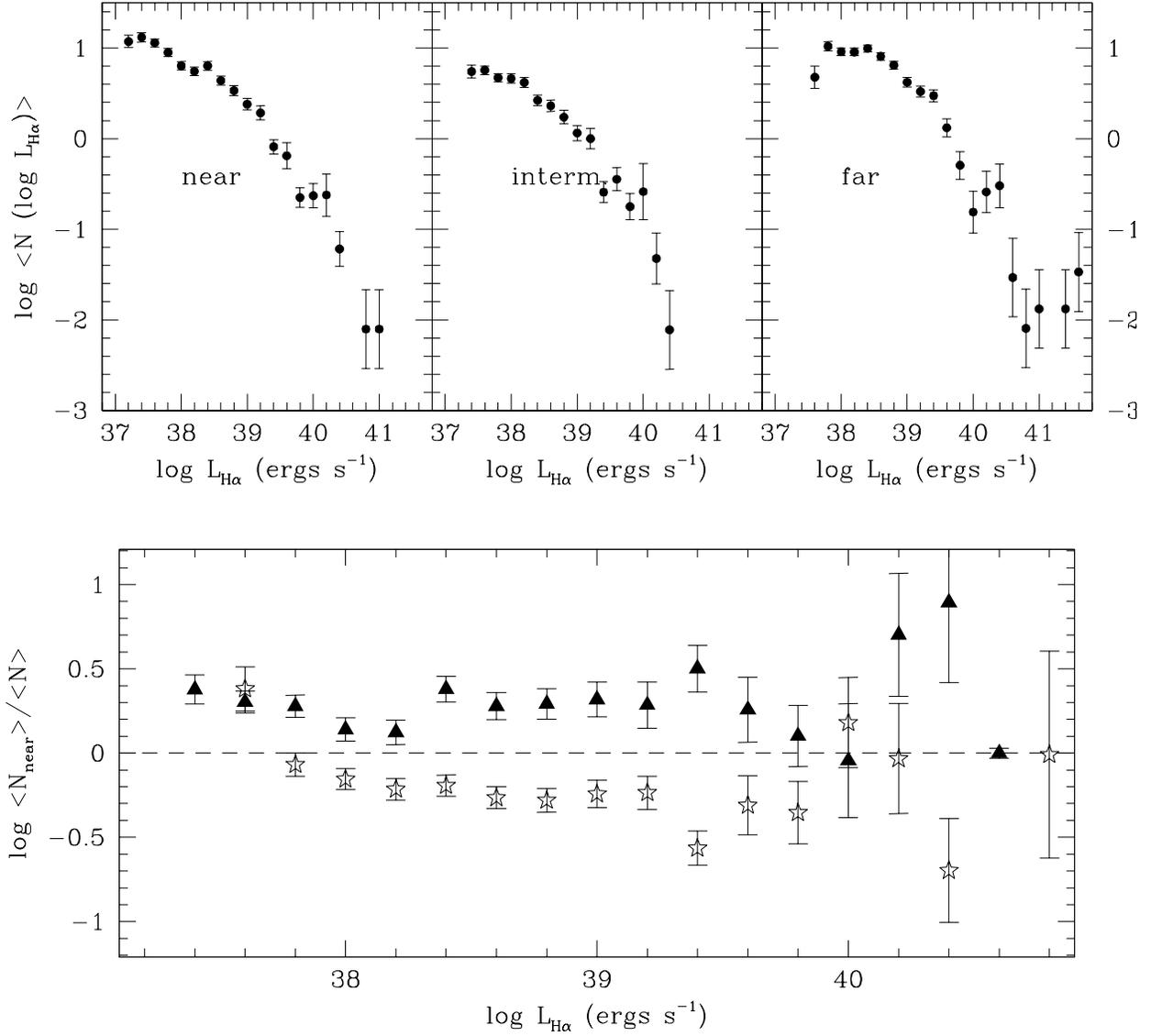}
\caption{Co-added HII region LFs for three different distance bins (upper); the ratio of the LF for the near group to that of the intermediate group (closed triangles, lower) and the ratio of the LF for the near group to that of the far group (stars, lower).}
\label{fig8}
\end{figure}

\begin{figure}
\plotone{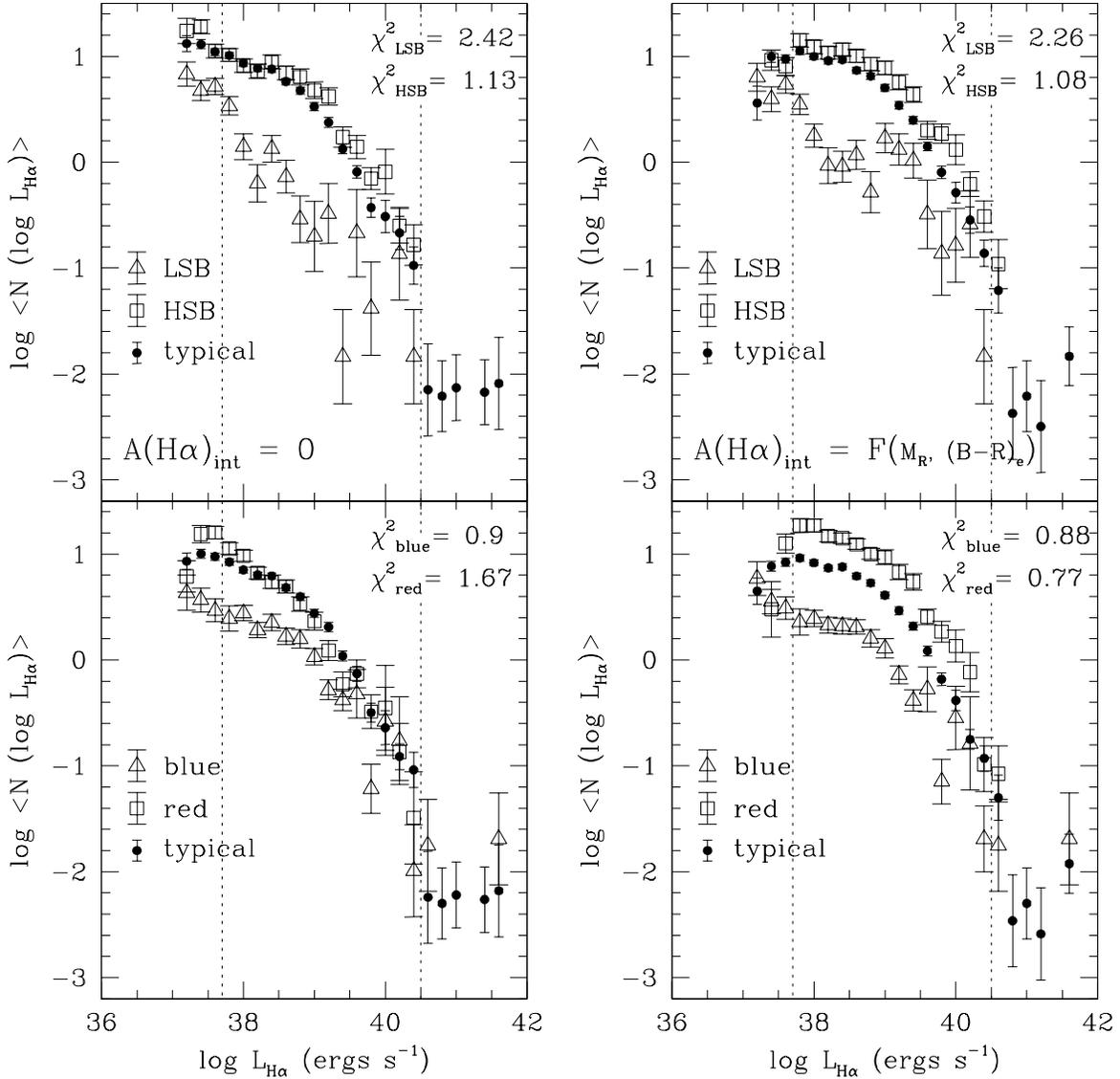}
\caption{Upper left: the co-added HII region LFs for the LSB (open triangles), HSB (open boxes), and surface brightness comparison (points) group LFs (see \S 3.2.2) with values for the reduced $\chi^{2}$ between the comparison LF and both the LSB and HSB group LFs (see \S 3.2.2).  Upper right:  same as upper left, but with internal extinction corrections applied according to equation (4).  Lower left: the co-added HII region LFs for the blue (open triangles), red (open boxes), and color comparison (points) group LFs (see \S 3.2.2) with values for the reduced $\chi^{2}$ between the comparison LF and both the blue and red group LFs (see \S 3.2.2).  Lower right:  same as lower left, but with internal extinction corrections applied according to equation (4).  In all panels, the range in log L$_{H\alpha}$ over which $\chi^{2}$ was calculated is marked with vertical dotted lines.}
\label{fig9}
\end{figure}

\placefigure{fig8}
\placefigure{fig9}

\subsection{Surface Brightness and Blending}
Using narrow-band images obtained with HST, \citet{sco01} found that in ground-based images of galaxies as nearby as M51 ($d \approx$10 Mpc), a significant fraction of H{\sc ii} regions will be blended, causing the measured H{\sc ii} region LF to be artificially flattened.  They argue that blending will have a larger effect on LFs measured for locations where the number density of H{\sc ii} regions is higher.  For example, they assert that this could explain why the LFs for interarm regions appear steeper than those for regions found in in spiral arms.  While Fig. 8 indicates that the effects of blending on the measured LFs do not change with distance within the relatively small volume covered by our sample, the effects of blending may be worse in higher surface brightness galaxies where the density of H{\sc ii} regions is likely to be higher, just as the density of H{\sc ii} regions is higher in spiral arms.  To explore this, we computed the projected distance to the nearest neighboring region for each H{\sc ii} region with {\sc snr}$>$5 within each galaxy which we will refer to as $\Delta_{min}$.  For the LSB, comparison, and HSB co-added groups, the median values for $\Delta_{min}$ are 420, 380, and 300 pc respectively, confirming that on average, the density of H{\sc ii} regions is higher in higher surface brightness galaxies.  Since we are limited by the resolution of our ground-based images and the lack of higher resolution images for the galaxies in our sample, we cannot explicitly determine the effect blending may have on the shape of the LF for each co-added group.  However, the shapes of the LFs for the HSB and comparison groups agree quite well even though the typical densities of H{\sc ii} regions for the two groups are significantly different.  It is therefore unlikely that the relatively lower densities of regions within the galaxies of the LSB group contribute significantly to the comparatively steeper shape of the LSB group LF.

\subsection{Dust Extinction}
The discrepancy in LF shapes may be influenced by the fact that no correction was made for internal extinction, especially in the case of the red group LF.  To explore this, we applied a correction for internal extinction at H$\alpha$, $A(H\alpha)_{int}$, derived from a linear least squares fit to the data of \citet{jan00} given by
\begin{equation}
\mbox{log } A(H\alpha)_{int} = (-0.063\pm0.021)M_{R} + (1.33\pm0.16)(B-R)_{e} + (-3.08\pm0.40)
\end{equation}
where $A(H\alpha)_{int}$ was derived using the measured Balmer decrements taken from the \citet{jan00} data, an intrinsic ratio of H$\alpha$ to H$\beta$ flux of 2.85 \citep{ost89}, the extinction curve of \citet{odo94}, and $R_{V}$=3.1.  As with the $[$NII$]$ correction given by equation (1), fitting a plane to the $A(H\alpha)_{int}$ data as a function of both $M_{R}$ and $(B-R)_{e}$ yielded a relation with $\sim$0.1 dex less scatter than the relation between $A(H\alpha)_{int}$ and $M_{R}$ used in paper I.  Such a correction may be inappropriate for studying the properties of the brightest H{\sc ii} regions as the extinction can vary significantly from one location to the next within a galaxy \citep{ken88}.  However, for the purpose of constructing co-added LFs, such a statistical correction provides a reasonable estimate of the effect that attenuation by dust has on the measured LFs.\par
HII{\it phot} was rerun on the images after the extinction corrections were applied.  We chose to rerun HII{\sc phot} rather than simply apply the extinction corrections to the measured H{\sc ii} region luminosities because the region boundaries are defined by the slope of their surface brightness profiles (see \S 2.2) which will be altered by the multiplicative extinction corrections.  However, we note that this is only relevant for relatively isolated regions.  The value for the limiting H{\sc ii} region luminosity was redetermined for each galaxy using the new HII{\it phot} output, and the co-added LFs were reconstructed following the procedure described above.  These LFs are plotted in the two right panels in Fig. 9.  For the new co-added LFs, the values for $\chi^{2}_{red}$ were recomputed as described above and are printed in the appropriate panels in Fig. 9.  From these plots and the $\chi^{2}_{red}$ values, it can be seen that using the extinction corrections causes the number of H{\sc ii} regions per galaxy to increase for both the HSB and red groups and that the shapes of the LFs for all of the color groups agree within less than 1$\sigma$.  For both the LSB and HSB group LFs, the values for $\chi^{2}_{red}$ are slightly smaller than they were for the LFs that did not use extinction corrections.  However, the LSB group LF has a relatively irregular shape and its value for $\chi^{2}_{red}$ is still relatively large at about 2.3.

\subsection{Small Number Statistics}
Since the number of regions with {\sc snr}$>$5 is less than 60 for all but one of the ten galaxies in the LSB group and is as low as four, the fact that the co-added LF for that group differs in shape from the LFs for the other groups may be the result of small number statistics.  To explore this possibility, we reconstruct the LF for each of the galaxies in the LSB group in the following way.  We first assume that the LF for the comparison surface brightness group is the ``true'' LF over the interval 37.7$\leq$log L$_{H\alpha} \leq$40.5 (see \S 3.2.2).  For galaxies where the limiting H{\sc ii} region luminosity is more than 10$^{37.7}$ ergs s$^{-1}$, we consider only the interval log $L_{lim} \leq \mbox{log } L_{H\alpha} \leq$40.5.  For each galaxy, we counted the number of H{\sc ii} regions found within this interval, $N_{R}$, and generated a set of $N_{R}$ random numbers which we used to sample the assumed true LF.  Following this, we co-added the new LFs in the same manner as described above and computed a value of $\chi^{2}_{red}$ between each fake co-added LF and the measured LSB group LF.  We repeated this procedure 1000 times to explore the range in shapes that could be produced by randomly sampling the assumed true LF with small numbers of regions.  As a control, we also perform the same procedure for the galaxies in the HSB group which have a median number of regions with {\sc snr}$>$5 of $\sim$150.\par
The first five fake co-added LFs created with this procedure using the data without internal extinction corrections are plotted in the upper panels of Fig. 10 with the actual LFs plotted in the upper left panel.  From these plots, it appears that the power law shape of the LSB group LF is most likely not the result of small number statistics while the shape of the HSB group LF is recovered in each instance.  We repeated the procedure using data with the internal extinction correction given in equation (4) applied to it; the first five fake co-added LFs are plotted in the lower panels of Fig. 10.  In this case, the LFs generated by randomly sampling the assumed true LF do appear to resemble the somewhat irregular shape observed for the LSB group.\par
To examine the results in a more quantitative manner, we plot the distribution of $\chi^{2}_{red}$ values in Fig. 11 for the LSB and HSB groups.  The results for the fake co-added LFs generated using data with no extinction corrections are plotted in the upper panel; the results for the fake co-added LFs generated using data with extinction corrections are plotted in the lower panel.  The modes for the $\chi^{2}_{red}$ distributions for the HSB group closely match the values printed in the upper panels of Fig. 9 which is what one would expect if the LFs for the galaxies in the HSB group are approximately as well sampled as the LFs for the galaxies in the comparison group.  The mode for the $\chi^{2}_{red}$ distribution for the LSB group without internal extinction corrections is about 1.9.  This is significantly lower that the value printed in Fig. 9 but is still not as low as that computed for the HSB group.  The mode for the $\chi^{2}_{red}$ distribution for the LSB group with extinction corrections is about 1.1 and the shape of the distribution is similar to that for the HSB group.  However, the mode is still larger than that found for the HSB group.\par
We also note that while the sample of \citet{jan00} that was used to derive the extinction corrections does span the same range in R-band luminosity as our sample, it contains a substantially lower fraction of LSB galaxies.  Since LSB galaxies have been found to have on average relatively low metallicities \citep[about 1/3 solar;][]{mcg94} and typically contain little if any molecular gas \citep{one03}, it is likely that equation (4) overestimates the amout of internal extinction for the LSB galaxies in our sample.  Taking this into account, the results displayed in Fig. 11 imply that while the effects of dust extinction and small number statistics contribute significantly to the shapes of the co-added LFs, they cannot fully explain the discrepancy between the shapes of the LSB and comparison group co-added LFs displayed in Fig. 9.  These results also imply that while small number statistics may strongly influence the trend found between $\mu_{e,R}$ and L$_{3}$ discussed in \S 2.5, there is also a real correlation between surface brightness and the luminosity of the brightest H{\sc ii} regions.

\begin{figure}
\plotone{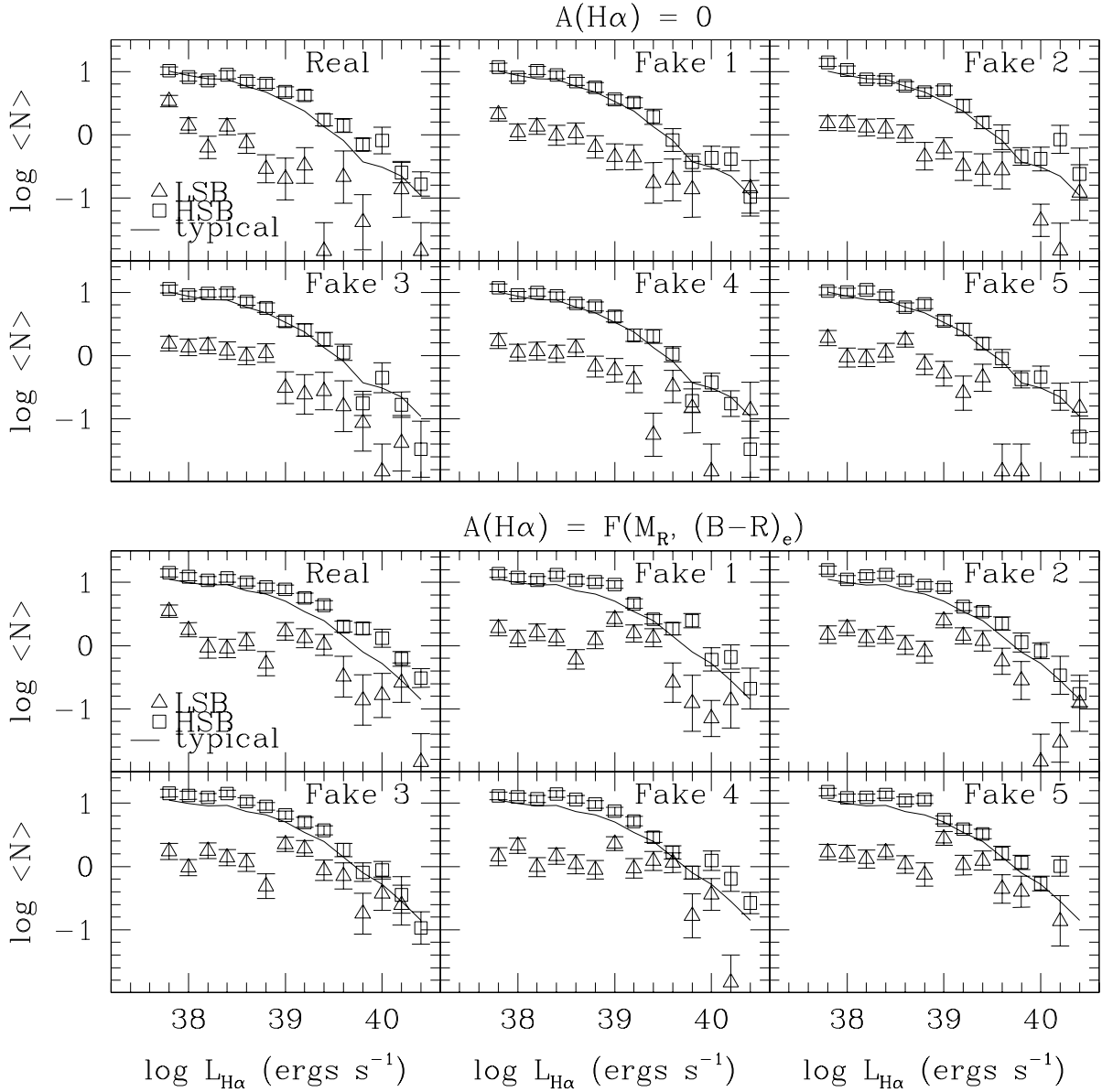}
\caption{Co-added HII region LFs generated for the LSB (open triangles) and HSB (open boxes) galaxy groups assuming that they randomly sample the same LF as is measured for the comparison surface brightness group (solid line).  The upper panels display results where no correction was applied for internal extinction; the lower panels display results for LFs that include the extinction corrections given by equation (4).  In each of the two sets of plots, the LFs in the upper left panel are the actual LFs measured from the data; the remaining panels represent the results of using the first five of 1000 different sets of random numbers to sample the assumed true LF (see \S 4).}
\label{fig10}
\end{figure}

\begin{figure}
\plotone{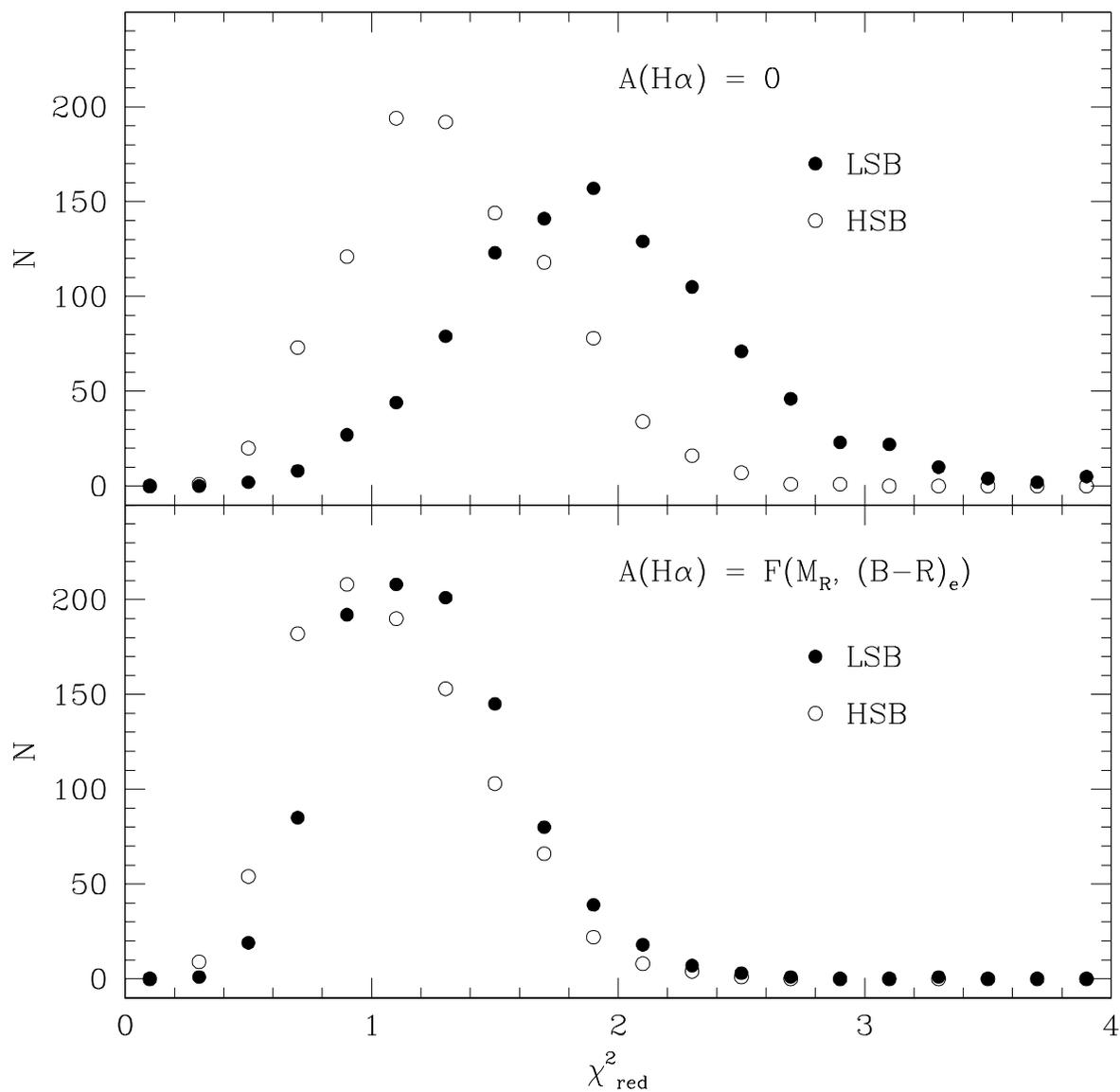}
\caption{Distributions for the reduced $\chi^{2}$ between the LSB (solid points) and HSB (open points) group co-added LFs and the 1000 fake LFs generated by randomly sampling the comparison group LF.  The upper panel displays the results for LFs that do not include the extinction corrections given by equation (4); the lower panel shows the results for LFs that do include extinction corrections.}
\label{fig11}
\end{figure}

\placefigure{fig10}
\placefigure{fig11}

\section{Discussion and Conclusions}
The trends displayed in Fig. 6 as well as the co-added LFs plotted in Fig. 9 suggest that the conditions under which star formation takes place in lower surface brightness galaxies may be different than in higher surface brightness systems.  This possibility is of particular interest since optically selected catalogs tend to be biased against these galaxies.  This observational bias is particularly relevant in regards to the shape of the H{\sc ii} region LF since eight out of the ten galaxies contained within the LSB co-added group have values for $\mu_{e,B}$ that are more than 2$\sigma$ fainter than the mean for the NFGS.  Since the extinction corrections derived from the \citet{jan00} data most likely overestimate the amount of dust in our LSB galaxies, the results in Fig. 11 imply that the difference in LF shapes can only be partially explained with a combination of the effects of dust extinction and small number statistics.  The effects of blending which may be stronger for higher surface galaxies do not appear to significantly contribute to the difference between the shapes of the comparison and LSB group LFs.  Therefore, this difference most likely reflects a real discrepancy between the distribution of H{\sc ii} region luminosities in LSB galaxies and that for relatively higher surface brightness spiral galaxies.\par
In terms of gas density and surface brightness, the disks of LSB galaxies are more similar to the extreme outer disks of more typical spiral galaxies than they are to the entire disk components of those galaxies.  It is therefore reasonable to compare the properties of the H{\sc ii} regions that have been found in the outer portions of spiral disks \citep[e.g][]{fer98, lel00, thi05} to those that we have observed for the LSB galaxies in our sample.  The luminosities found for the star forming regions in the outer disks of NGC 628 \citep{lel00} and M83 \citep{thi05} imply that the H$\alpha$ luminosities for the vast majority of such regions are less than $\sim10^{38}$ ergs s$^{-1}$.  The LF for the LSB group indicates that LSB galaxies are capable of forming a significant number of regions with L$_{H\alpha}>10^{38}$ ergs s$^{-1}$ and in some cases, a few regions with L$_{H\alpha}>10^{40}$ ergs s$^{-1}$.  This implies that while H{\sc ii} regions in LSB galaxies are less luminous than those in higher surface brightness spiral galaxies, LSB galaxies may be capable of forming more luminous regions than those found in the outer portions of nearby disk galaxies.  However, we note that the number of galaxies for which outer disk H{\sc ii} regions have been study in detail is small and that future observations may reveal that it is possible for larger H{\sc ii} regions to form in these locations as well.\par
The difference in shape between the LSB group and comparison LFs displayed in the upper left panel of Fig. 9 resembles the observed difference between arm and interarm H{\sc ii} region LFs \citep[e.g.][]{ken89, ban93, oey98, thi00} with the LSB group LF more closely resembling the typical interarm H{\sc ii} region LF.  The fact that interarm H{\sc ii} regions are on average less luminous that those contained in spiral arms is consistent with the idea that the star clusters that are ionizing the regions within spiral arms are younger that those that are ionizing interarm regions \citep{oey98}.  \citet{sco01} have argued that the difference in LF shapes may result from the effects of blending being less severe for interarm regions where the typical spacing between H{\sc ii} regions is relatively larger.  However, as discussed in \S 3.2, the difference in shape between the LF for the LSB group, where the spacing between H{\sc ii} regions is typically larger, and that for the comparison group LF is most likely not the result of the comparison group LF being flattened more by blending than the LSB group LF.\par
The fact that the LSB group LF is similar to interarm region LFs may then imply that the star formation in LSB galaxies is somewhat episodic so that on average, the relatively new star clusters found in these galaxies will be slightly older than those found in typical spirals that are continually forming stars.  Indeed, the relatively blue B-H colors of many gas rich, LSB galaxies \citep[see ][]{bot84, deb95} cannot be reconciled with an exponentially declining or constant star formation history (over the last few billion years).  Such star formation histories would cause the giant branch to be more populated yielding redder colors.  To simultaneously explain the blue broad band colors and low abundances in systems which do form massive stars as evidenced by the LSB group co-added H{\sc ii} region LF, it seems likely that episodic star formation events define the recent star formation history of the typical LSB disk.\par
The difference in LF shape along with the trend between surface brightness and L$_{3}$ seen in Fig. 6 may also result from LSB galaxies forming a higher fraction of stars in lower mass star clusters or outside of clusters altogether.  Indeed, an interesting observation yet to be performed is to examine whether the frequency of stellar clusters in LSB disks is similar to that seen in normal disk galaxies.  The toy model of \citet{one98} argues that stellar cluster formation in LSB disk galaxies is suppressed (because of the very large Jeans length in these low density disks) and that the formation of individual massive stars occurs stochastically within the disk and not exclusively within stellar clusters.  While there is yet no direct observational evidence for a paucity of stellar clusters in LSB disks, our data is at least indirectly consistent with this view.  The trend between surface brightness and diffuse fraction may support this notion as \citet{hoo00} estimates that main sequence O and B stars that reside in the field rather than in H{\sc ii} regions account for $\sim$40\% of the ionization of the DIG in typical spiral galaxies.  If a relatively larger fraction of the O and B stars in LSB galaxies are not formed in dense, massive molecular clouds that would produce H{\sc ii} regions, they would be expected to have larger diffuse fractions as observed.  In any case, the results imply that the manner in which star formation proceeds in LSB galaxies is considerably different than that for more HSB spiral galaxies.\par
The authors would like to thank the NOAO TAC for allocation of observing time and the CTIO staff for expert assistance at the telescope.  We would also like to thank D. Thilker for assisting with the usage of HII{\it phot} and the referee for useful comments and suggestions.

\clearpage

\clearpage

\clearpage

\clearpage

\clearpage

\clearpage

\clearpage

\clearpage

\clearpage

\clearpage

\clearpage

\clearpage

\begin{deluxetable}{lrrrrrrrr}
\tablecolumns{9}
\tablewidth{0pc}
\tablecaption{DIG and HII Region Properties}
\tablehead{
\colhead{} & \colhead{} & \colhead{} & \colhead{} & \colhead{} & \colhead{log L$_{3}$} & \colhead{D$_{3,corr}$} & \colhead{} & \colhead{d} \\
\colhead{Galaxy} & \colhead{N(HII)$_{38}$} & \colhead{log $\frac{N(HII)_{38}}{L_{R}}$} & \colhead{$f_{d}$} & \colhead{$\sigma_{f_{d}}$} & \colhead{(ergs s$^{-1}$)} & \colhead{(pc)} & \colhead{$\alpha$} & \colhead{(Mpc)} \\
\colhead{(1)} & \colhead{(2)} & \colhead{(3)} & \colhead{(4)} & \colhead{(5)} & \colhead{(6)} & \colhead{(7)} & \colhead{(8)} & \colhead{(9)}}

\startdata
         ESO004-G017 &     3 &  -6.76 & 0.83 & 0.06 &  38.5 &    406 & $\cdots$ &  25.0 \\
         ESO013-G016\tablenotemark{a} &    58 &  -6.16 & 0.42 & 0.10 &  39.4 &    753 & $\cdots$ &  24.8 \\
         ESO017-G002\tablenotemark{a} &    13 &  -6.72 & 0.18 & 0.13 &  39.3 &    583 & $\cdots$ &  23.0 \\
         ESO019-G004 &     6 &  -6.57 & 0.70 & 0.08 &  38.6 &    207 & $\cdots$ &  28.8 \\
         ESO021-G003 &   109 &  -6.26 & 0.20 & 0.19 &  40.1 &    327 &  -1.9 &  32.7 \\
         ESO027-G001\tablenotemark{a} &   185 &  -6.50 & 0.60 & 0.04 &  40.2 &    689 &  -1.8 &  36.3 \\
         ESO027-G021\tablenotemark{a} &    39 &  -6.42 & 0.43 & 0.05 &  39.8 &    678 &  -1.2 &  34.8 \\
         ESO035-G009\tablenotemark{a}$\;\;$\tablenotemark{b} &     3 &  -6.44 & 0.73 & 0.11 &  38.1 &    330 & $\cdots$ &  15.7 \\
         ESO035-G018\tablenotemark{a}$\;\;$\tablenotemark{b} &    63 &  -6.35 & 0.55 & 0.06 &  39.4 &    781 & $\cdots$ &  24.9 \\
         ESO035-G020 &    21 &  -6.23 & 0.33 & 0.01 &  39.7 &   1050 & $\cdots$ &  31.6 \\
         ESO035-G021 &     7 &  -6.84 & 0.64 & 0.13 &  39.2 &    213 & $\cdots$ &  17.4 \\
         ESO036-G006 &    17 &  -6.26 & 0.67 & 0.17 &  38.7 &    236 & $\cdots$ &  15.9 \\
         ESO037-G004 &     2 &  -6.88 & 0.70 & 0.07 &  38.2 &    383 & $\cdots$ &  18.1 \\
         ESO037-G010 &   233 &  -6.30 & 0.27 & 0.02 &  39.9 &    179 &  -1.8 &  25.4 \\
         ESO037-G015\tablenotemark{a} &    23 &  -5.92 & 0.39 & 0.03 &  39.0 &    543 & $\cdots$ &  23.0 \\
         ESO038-G011 &    13 &  -6.66 & 0.47 & 0.05 &  39.2 &    966 & $\cdots$ &  27.0 \\
         ESO054-G021\tablenotemark{a} &    64 &  -6.25 & 0.63 & 0.03 &  39.6 &    945 & $\cdots$ &  19.7 \\
         ESO059-G001 &     2 &  -6.57 & 0.43 & 0.06 &  38.1 &    315 & $\cdots$ &   6.4 \\
         ESO060-G007\tablenotemark{a} &     5 &  -6.17 & 0.78 & 0.06 &  38.4 &    169 & $\cdots$ &  21.5 \\
         ESO060-G019 &    57 &  -6.50 & 0.20 & 0.05 &  40.0 &    894 &  -1.5 &  20.4 \\
        ESO060-IG003 &    10 &  -6.15 & 0.26 & 0.03 &  39.0 &    217 & $\cdots$ &  20.0 \\
         ESO061-G017 &    13 &  -6.37 & 0.41 & 0.10 &  38.8 &    226 & $\cdots$ &  25.2 \\
         ESO079-G005\tablenotemark{a}$\;\;$\tablenotemark{b} &    25 &  -6.23 & 0.45 & 0.03 &  39.4 &    495 & $\cdots$ &  24.1 \\
         ESO079-G007\tablenotemark{a}$\;\;$\tablenotemark{b} &    39 &  -6.12 & 0.49 & 0.14 &  39.4 &    175 & $\cdots$ &  23.3 \\
         ESO080-G005\tablenotemark{a}$\;\;$\tablenotemark{b} &    14 &  -6.38 & 0.46 & 0.06 &  39.2 &    742 & $\cdots$ &  21.2 \\
         ESO084-G040 &     2 &  -6.51 & 0.32 & 0.07 &  38.4 &    306 & $\cdots$ &  16.6 \\
         ESO085-G014 &    43 &  -6.08 & 0.41 & 0.02 &  39.6 &    712 &  -1.7 &  19.0 \\
         ESO085-G030\tablenotemark{a} &    23 &  -6.01 & 0.23 & 0.05 &  39.4 &    359 & $\cdots$ &  17.5 \\
         ESO085-G047 &    23 &  -5.97 & 0.49 & 0.05 &  39.2 &    594 & $\cdots$ &  19.8 \\
         ESO086-G060 &     3 &  -6.23 & 0.74 & 0.06 &  38.2 &    151 & $\cdots$ &  22.6 \\
         ESO090-G004 &    35 &  -6.11 & 0.56 & 0.08 &  39.3 &    270 & $\cdots$ &  29.8 \\
         ESO091-G007 &    42 &  -6.61 & $\cdots$ & $\cdots$ &  39.8 &    400 & $\cdots$ &  31.1 \\
         ESO092-G006 &   203 &  -6.24 & 0.11 & 0.30 &  40.1 &    756 &  -1.8 &  24.9 \\
         ESO092-G021 &   433 &  -6.06 & 0.46 & 0.05 &  40.3 &   1430 &  -1.9 &  29.1 \\
         ESO104-G022 &     7 &  -6.18 & 0.64 & 0.07 &  38.9 &    439 & $\cdots$ &  11.5 \\
         ESO119-G016\tablenotemark{a} &     5 &  -6.11 & 0.60 & 0.23 &  38.5 &    376 & $\cdots$ &  12.6 \\
         ESO140-G019 &     1 &  -6.44 & 0.12 & 0.02 &  38.2 &    616 & $\cdots$ &  13.9 \\
       HIPASS0635-70\tablenotemark{a} & $\cdots$ & $\cdots$ & $\cdots$ & $\cdots$ & $\cdots$ & $\cdots$ &  22.6 \\
       HIPASS0653-73 &     7 &  -5.51 & $\cdots$ & $\cdots$ &  39.2 & $\cdots$ & $\cdots$ &  16.7 \\
       HIPASS1039-71 &     2 &  -6.07 & $\cdots$ & $\cdots$ &  37.9 &     50 & $\cdots$ &  22.0 \\
              IC2150\tablenotemark{a} &    84 &  -6.62 & 0.62 & 0.03 &  40.0 &   1057 &  -1.2 &  24.4 \\
              IC2554\tablenotemark{a}$\;\;$\tablenotemark{b} &    17 &  -6.23 & 0.46 & 0.02 &  39.8 &    559 &  -1.3 &  19.8 \\
              IC3104 &     5 &  -6.82 & 0.64 & 0.23 &  38.9 &    106 & $\cdots$ &   5.9 \\
              IC4662 &    63 &  -5.69 & 0.03 & 0.12 &  39.9 &    172 & $\cdots$ &   5.6 \\
              IC4710\tablenotemark{a} &    31 &  -6.14 & 0.55 & 0.04 &  39.2 &    349 & $\cdots$ &  10.6 \\
              IC4870\tablenotemark{a} &    10 &  -5.43 & 0.19 & 0.19 &  39.2 &    277 & $\cdots$ &  12.6 \\
              IC5028\tablenotemark{a}$\;\;$\tablenotemark{b} &    35 &  -5.49 & 0.67 & 0.05 &  39.2 &    218 & $\cdots$ &  23.6 \\
              IC5176\tablenotemark{a}$\;\;$\tablenotemark{b} &    57 &  -6.37 & 0.71 & 0.09 &  39.1 &    595 & $\cdots$ &  24.9 \\
                 M83 &  2340 &  -5.79 & $\cdots$ & $\cdots$ &  40.6 &    406 & $\cdots$ &  10.7 \\
             NGC0406\tablenotemark{a} &    55 &  -6.21 & 0.53 & 0.03 &  39.8 &    472 &  -1.6 &  21.1 \\
             NGC0802\tablenotemark{a} &     2 &  -7.05 & 0.33 & 0.07 &  38.4 &   1214 & $\cdots$ &  20.4 \\
             NGC1313 &   160 &  -5.67 & 0.10 & 0.03 &  39.5 &    272 & $\cdots$ &   5.0 \\
             NGC1511\tablenotemark{a} &    87 &  -6.33 & 0.18 & 0.05 &  40.4 &    591 &  -1.5 &  18.2 \\
             NGC1559\tablenotemark{a} &   259 &  -6.13 & 0.27 & 0.03 &  40.3 &    755 &  -1.7 &  17.5 \\
             NGC1809\tablenotemark{a} &   153 &  -6.34 & 0.43 & 0.83 &  39.5 &    274 & $\cdots$ &  17.8 \\
             NGC1892 &    52 &  -6.24 & 0.37 & 0.02 &  39.8 &    667 &  -1.7 &  18.4 \\
             NGC2082\tablenotemark{a} &    66 &  -6.03 & 0.20 & 0.13 &  39.4 &    484 & $\cdots$ &  16.1 \\
            NGC2397B &     9 &  -6.30 & 0.37 & 0.04 &  38.9 &    144 & $\cdots$ &  19.8 \\
             NGC2442\tablenotemark{a} &   278 &  -6.56 & 0.45 & 0.05 &  39.6 &    663 & $\cdots$ &  20.8 \\
            NGC2788B\tablenotemark{a} &    11 &  -6.31 & 0.48 & 0.03 &  38.8 &    478 & $\cdots$ &  20.1 \\
             NGC2836 &    74 &  -6.34 & 0.55 & 0.18 &  39.5 &    660 & $\cdots$ &  24.1 \\
             NGC2915 &     8 &  -6.22 & 0.16 & 0.06 &  39.0 &    242 & $\cdots$ &   5.8 \\
             NGC3059 &   433 &  -6.05 & 0.18 & 0.02 &  40.5 &    228 &  -1.9 &  17.6 \\
            NGC3136A &    92 &  -5.64 & 0.38 & 0.14 &  40.0 &    106 &  -1.6 &  28.9 \\
             NGC5068\tablenotemark{a} &   396 &  -5.89 & 0.22 & 0.05 &  40.1 &    913 &  -2.0 &  12.6 \\
             NGC6300 &   344 &  -6.34 & 0.23 & 0.63 &  40.1 &     77 &  -2.3 &  16.0 \\
            NGC6438A &   176 &  -6.90 & 0.39 & 0.08 &  40.4 &    889 &  -1.6 &  36.4 \\
             NGC7098\tablenotemark{a} &    97 &  -6.97 & 0.90 & 0.07 &  39.4 &    587 & $\cdots$ &  33.6 \\
             NGC7661\tablenotemark{a} &    35 &  -6.15 & 0.71 & 0.12 &  39.0 &    496 & $\cdots$ &  28.9 \\
\enddata

\tablecomments{Col. (2): The number of H{\sc ii} regions found by HII{\it phot} with L$_{H\alpha} > 10^{38}$ ergs s$^{-1}$ cm$^{-2}$.  Col. (3): Number of H{\sc ii} regions with L$_{H\alpha} > 10^{38}$ ergs s$^{-1}$ cm$^{-2}$ relative to the R-band luminosity (in solar units).  Col. (4): The fraction of the total H$\alpha$ flux attributed to the DIG (the diffuse fraction).  Col. (5): The 1$\sigma$ error in the diffuse fraction.  Col. (6): The weighted mean H$\alpha$ luminosity of the three brightest H{\sc ii} regions.  Col. (7): The weighted mean diameter of the three brightest H{\sc ii} regions corrected for the effects of the {\sc psf}.  Col. (8): The slope of the H{\sc ii} region luminosity function for log L$_{H\alpha}>$39.  Col. (9): The distance to the galaxy for H$_{\circ}$=70 km s$^{-1}$ Mpc$^{-1}$.}
\tablenotetext{a}{The images for this galaxy were most likely taken through clouds, and the values listed in Col. (3) and (6) should be taken as estimates only}
\tablenotetext{b}{The H$\alpha$ flux calibration for this galaxy was derived using the calibration for the R-band (see paper I).}
\label{tab1}
\end{deluxetable}

\end{document}